\documentclass[journal]{IEEEtran}
\usepackage{graphicx}
\usepackage{epsfig}
\usepackage{multirow}
\usepackage{subfigure}
\usepackage{amsmath,amssymb}
\usepackage[linesnumbered,ruled]{algorithm2e}
\usepackage{cite}
\usepackage{balance}
\usepackage{epsfig}
\usepackage{graphicx}
\usepackage{color}
\usepackage[english]{babel}
\usepackage{amsfonts,amsthm,amsmath}
\usepackage{times}
\usepackage{multirow}
\usepackage{subfigure}
\usepackage{amsmath}
\usepackage{amsfonts}
\usepackage{psfrag}
\graphicspath{{./paper_figures/}}

\usepackage[linesnumbered,ruled]{algorithm2e}

\theoremstyle{plain}

\newtheorem{remark}{Remark}

\begin{document}

\title{Microphone Subset Selection for MVDR Beamformer Based Noise Reduction}

\author{Jie Zhang, Sundeep Prabhakar Chepuri, Richard C. Hendriks, and Richard Heusdens
\thanks{
Manuscript received xxxxx xx, 2016; revised xxxxx xx, 2016; accepted
xxxxx xx, 2016. Date of publication xxxxx xx, 2016; date of current version
xxxxx xx, 2016. This work is supported by the China Scholarship Council (NO. 201506010331). The associate editor coordinating the review of this manuscript and approving
it for publication was Prof. ********.
The authors are with the Faculty of Electrical Engineering, Mathematics and
Computer Science, Delft University of Technology, 2628 CD Delft, The Netherlands (e-mail:
\{j.zhang-7, s.p.chepuri, r.c.hendriks, r.heusdens\}@tudelft.nl).}
\thanks{Color versions of one or more of the figures in this paper are available online
at http://ieeexplore.ieee.org.}
\thanks{Digital Object Identifier: **************}
}
\markboth{IEEE/ACM Transactions on Acoustics, Speech and Language Processing,~Vol.~**, No.~*, **~2017}%
{Shell \MakeLowercase{\textit{et al.}}: Bare Demo of IEEEtran.cls for IEEE Journals}

\maketitle
\begin{abstract}
In large-scale wireless acoustic sensor networks (WASNs), many of the sensors will only have a marginal contribution to a certain estimation task. Involving all sensors increases the energy budget unnecessarily and decreases the lifetime of  the WASN. Using microphone subset selection, also termed as sensor selection, the most informative sensors can be chosen from a set of candidate sensors to achieve a prescribed inference performance.
In this paper, we consider microphone subset selection for minimum variance distortionless response (MVDR) beamformer based noise reduction. The best subset of sensors is determined by minimizing the transmission cost while constraining the output noise power (or signal-to-noise ratio). Assuming the statistical information on correlation matrices of the sensor measurements is available, the sensor selection problem for this model-driven scheme is first solved by utilizing convex optimization techniques. In addition, to avoid estimating the statistics related to all the candidate sensors beforehand, we also propose a data-driven approach to select the best subset using a greedy strategy. The performance of the greedy algorithm converges to that of the model-driven method, while it displays advantages in dynamic scenarios as well as on computational complexity. Compared to a sparse MVDR or radius-based beamformer, experiments show that the proposed methods can guarantee the desired performance with significantly less transmission costs.
\end{abstract}
\IEEEpeerreviewmaketitle

\section{Introduction}\label{sec:intro}
\IEEEPARstart{M}{icrophone} arrays have become increasingly popular in many speech processing applications, e.g., hearing aids [1], teleconferencing systems [2], hands-free telephony [3], speech recognition [4], human-robot interaction [5], etc. Compared to their single-microphone counterparts, microphone arrays typically lead to an enhanced performance when detecting, localizing, or enhancing specific sound sources. This is due to the fact that with a microphone array the sound field is  not only sampled in time, but also in space.

Although traditional microphone arrays have been widely investigated, see [6] and reference therein, they do have some important limitations. Typically, conventional microphone arrays have one central  processing unit, that is, a fusion center (FC), which physically connects to the microphones. Rearranging the microphones in such a conventional  wired and centralized array is impractical. Moreover, usually the target source is located far away from the array, resulting in a low signal-to-noise ratio (SNR). In addition, typically, the size of conventional arrays is limited as the maximum array size is determined by the application device [7].

Recently, wireless acoustic sensor networks (WASNs) have attracted an increased amount of interest [7]-[10]. In a WASN, each sensor node is equipped with a single microphone or a small microphone array, and the nodes are spatially distributed across a specific environment. The microphone nodes communicate with their neighboring
nodes or the FC using wireless links.
The use of WASNs can potentially resolve the limitations encountered with the conventional arrays that were mentioned before. At first, the WASN is not constrained to any specific (fixed) array configuration. Secondly, with a WASN, the position and number of microphones is not anymore determined by the application device. Instead,  microphones can be placed at positions that are difficult to reach with conventional microphones.  With a WASN, the array-size limitations disappear and the network becomes scalable (i.e., larger array apertures can be achieved) [11]. The fact that microphones in the WASN sample the sound field in a much larger area can yield higher quality recordings as it is likely that some of the sensors are close to the target source and have a higher SNR. One of the bottlenecks in a WASN is the resource usage in terms of power. Transmission of data between nodes or from the nodes to the FC will influence the battery lifetime of the sensor. Although all microphones in the WASN will positively contribute to the estimation task,  only a few will have a significant contribution. It is questionable whether using all microphones in the network is beneficial taking the energy usage and lifetime of the sensors into account. Instead of blindly using all sensors, selecting a subset of microphones that is most informative for an estimation task at hand can reduce the data to be processed as well as transmission costs.

In this work, we investigate spatial filtering based noise reduction using only the most informative data via {\it{microphone subset selection}}, or so-called {\it{sensor selection}}, to reach a prescribed performance with low power consumption.
Sensor selection is important  for data dimensionality reduction. Mathematically, sensor selection is often expressed in terms of  the following optimization problem:
\begin{equation}
  \displaystyle \arg \min_{\mathbf{p}\in\{0,1\}^M}\ \ f(\mathbf{p})\ \ \ {\rm s. t.}\ \ \mathbf{1}_M^T\mathbf{p}=K,
\end{equation}
where $\mathbf{p}$ indicates whether a sensor is selected or not, and the cost function $f(\mathbf{p})$ is optimized to select the best subset of $K$ sensors out of $M$ available sensors. Basically, the problem in (1) is a non-convex Boolean optimization problem, which incurs a combinatorial search over all the $\binom{M}{K}$ possible combinations. Usually, it can be simplified via convex relaxation techniques [12]-[14] or using greedy heuristics, e.g., leveraging submodularity [15][16]. When the cardinality of $\mathbf{p}$ is of more concern, the cost function and constraint in (1) can also be interchanged by minimizing the cardinality of $\mathbf{p}$, i.e., $||\mathbf{p}||_0$, while constraining the performance meansure $f(\mathbf{p})$.

In general, sensor selection can be categorized into two classes: model-driven schemes and data-driven schemes. For the model-driven schemes, sensor selection is an offline design, where the sensing operation is designed based only on the data model (even before gathering data) such that a desired ensemble inference performance is achieved. In other words, the model-driven schemes provide the selected sensors {\it{a priori}} for the inference tasks [14]. There are many applications of the model-driven schemes for sensor placement in source localization [13], power grid monitoring [17], field estimation [18], target tracking [14], to list a few. In contrast to the offline design schemes, dimensionality reduction can also be done on already acquired data by discarding, i.e., censoring, less informative samples; this is referred as data-driven schemes. Data-driven sensor selection has been applied within the context of speech processing, e.g., speech enhancement [19][20], speech recognition [21], and target tracking by sensor scheduling [22]. In the WASNs context, due to time-varying topologies, we have typically no information about the data model (e.g., probability density function), but the online measured data (e.g., microphone recordings) are available instead. In this work, we start with the model-driven sensor selection for the spatial filtering based noise reduction problem, which is then extended to a data-driven scheme.


\subsection{Contributions}
In this paper, we consider the problem of selecting the most informative sensors for noise reduction based on the minimum variance distortionless response (MVDR) beamformer. We formulate this problem to minimize the total transmission power subject to a constraint on the performance. While the classical sensor selection problem formulation as also given in (1) puts a constraint on the number of selected sensors, in the speech enhancement context the desired number of sensors is typically unknown. Hence, the desired number of sensors heavily depends on the scenario, e.g., the number of  sound sources.
Within the speech enhancement context it would be more useful to relate the constraint to a certain performance in terms of the expected quality or intelligibility of the final estimated signal. We therefore reformulate the sensor selection problem to be constrained to a certain expected output performance. In such a way, the selected sensors are always the ones having the minimum transmission power.

The minimization problem is first solved by convex optimization techniques exploiting the available complete joint statistics (i.e., correlation matrices) of the microphone measurements of the complete network, such that the selected subset of microphones is optimal. This is referred as the proposed model-driven approach.

In a more practical scenario, usually it is impossible to estimate the joint statistics of the complete network  beforehand due to the dynamics of the scenario. Instead, the real-time measured data is only what can be accessed. Therefore, we extend the proposed model-driven algorithm to a data-driven scheme using a greedy sensor selection strategy. The performance of the greedy approach is proven to converge to that of the model-based method from an experimental perspective. There are a few existing  contributions considering microphone subset selection in the area of audio signal processing. For example, Szurley et al. greedily selected an informative subset according to the SNR gain at each individual microphone for speech enhancement [20]. Bertrand and Moonen [19] conducted greedy sensor selection based on the contribution of each sensor signal to mean squared error (MSE) cost for signal estimation. Kumatani et al. proposed a channel selection for distant speech recognition by considering the contribution of each channel to multichannel cross-correlation coefficients (MCCCs) [21]. The proposed greedy algorithm shows an advantage in  computational complexity and optimality as compared to existing greedy approaches [19][20].
\subsection{Outline and notation}
The rest of this paper is organized as follows. Sec.~\ref{sec:preliminary}  introduces the signal model, the classical MVDR beamforming, and sensor selection model. Sec.~\ref{sec:problem_formulation} presents the problem formulation. Sec.~\ref{sec:model-selection} presents two solvers based on convex optimization to solve the model-driven sensor selection problem.
Sec.~\ref{sec:greedy} proposes a greedy algorithm. Sec.~\ref{sec:exp} illustrates the simulation results. Finally, Sec.~\ref{sec:conclusion} concludes this work.

The notation used in this paper is as follows: Upper (lower) bold face letters are used for matrices (column vectors). $(\cdot)^T$ or $(\cdot)^H$ denotes (vector/matrix) transposition or conjugate transposition. $\text{diag}(\cdot)$ refers to a block diagonal matrix with the elements in its argument on the main diagonal. $\mathbf{1}_N$ and $\mathbf{0}_N$ denote the $N\times 1$ vector of ones and the $N\times N$ matrix with all its elements equal to zero, respectively. $\mathbf{I}_N$ is an identity matrix of size $N$. $\mathbf{A}\succeq \mathbf{B}$ means that $\mathbf{A}-\mathbf{B}$ is a positive semidefinite matrix. $|\mathcal{U}|$ denotes the cardinality of the set $\mathcal{U}$.

\section{Preliminaries}\label{sec:preliminary}
\subsection{Signal model}
We assume a spatially distributed candidate set of $M$ microphone sensors that collect and transmit their observations to an FC. The multi-microphone noise reduction methods considered in this paper operate in the frequency domain on
a frame-by-frame basis. Let $l$ denote the frame index and $\omega$ the frequency bin index, respectively. We assume that the user (i.e., FC) has one source of interest, while multiple interfering sources are present in the environment. Using a discrete Fourier transform (DFT) domain description, the noisy DFT coefficient at the $k$-th microphone, say $y_k(\omega, l)$, for $ k = 1,2,\cdots,M$, is given by
\begin{equation} \label{eq:signal_model_scale}
 y_k(\omega,l) = x_k(\omega,l) + n_k(\omega,l),
\end{equation}
where $x_k(\omega,l)=a_k(\omega)s(\omega,l)$ with $a_k(\omega)$ denoting the acoustic transfer function (ATF) of the target signal with respect to the $k$-th microphone and $s(\omega,l)$ the target source signal at the source location of interest. In (\ref{eq:signal_model_scale}), the component $n_k(\omega,l)$ represents the total received noise at the $k$-th microphone (including interfering sources and internal thermal additive noise).
For notational convenience, the frequency variable $\omega$ and the frame index $l$ will be omitted now onwards bearing in mind that the processing takes place in the frequency domain.
Using vector notation, signals from $M$ microphones are stacked in a vector $\mathbf{y} = [y_1,...,y_M]^T  \in \mathbb{C}^M$.
Similarly, we define an $M$  dimensional speech vector $\mathbf{x}$  for the speech component contained in $\mathbf{y}$ as $\mathbf{x} = \mathbf{a}s \in \mathbb{C}^M$ with $\mathbf{a}\in \mathbb{C}^M$, and length-$M$ noise vector ${\bf n}$. As a consequence, the signal model in (\ref{eq:signal_model_scale}) can be compactly written as
\begin{equation}
 \mathbf{y} = \mathbf{x} + \mathbf{n}.
\end{equation}

Assuming that the speech and noise components are mutually uncorrelated, the correlation matrix of the received signals is given by
\begin{equation}\label{eq:Ryy=Rxx+Rnn}
  \mathbf{R_{yy}} = \mathbb{E}\{\mathbf{y}\mathbf{y}^H\} = \mathbf{R_{xx}} + \mathbf{R_{nn}} \in \mathbb{C}^{M\times M},
\end{equation}
where $\mathbb{E}\{\cdot\}$ denotes the expectation operation, and $\mathbf{R_{xx}}=\mathbb{E}\{\mathbf{x}\mathbf{x}^H\}=P_s\mathbf{a}\mathbf{a}^H$ with $P_s=\mathbb{E}\{|s|^2\}$ representing the power spectral density (PSD) of the target source. Notice that due to the assumption that ${\bf{x}}$ and ${\bf{n}}$ are uncorrelated, $\mathbf{R}_{\mathbf{xx}}$ can be estimated by subtracting the noise correlation matrix $\mathbf{R}_{\mathbf{nn}}$, which is estimated during the absence of speech  from the  speech-plus-noise correlation matrix $\mathbf{R}_{\mathbf{yy}}$ [23].

\subsection{MVDR beamformer}\label{sec:mvdr}
The well-known MVDR beamformer minimizes the total output power after beamforming while simultaneously keeping the gain of the array towards the desired signal fixed. Therefore, any reduction in the output energy is obtained by suppressing interference or noise. Mathematically, this can be written as
\begin{equation}\label{eq:MVDR}
\begin{aligned}
 \hat{\mathbf{w}} = &\arg\min_{\mathbf{w}}\ \mathbf{w}^H\mathbf{R}_{\mathbf{nn}}\mathbf{w}, \\
   &\text{s. t. }\mathbf{w}^H\mathbf{a}=1.
\end{aligned}
\end{equation}
The optimal solution, in a best linear unbiased estimator sense, can be obtained using the method of Lagrange multipliers, and is given by [8][24][25]
\begin{equation}
  \hat{\mathbf{w}}=\frac{\mathbf{R}_{\mathbf{nn}}^{-1}\mathbf{a}}{\mathbf{a}^H\mathbf{R}_{\mathbf{nn}}^{-1}\mathbf{a}}.
\end{equation}

After processing by the MVDR beamformer, the output SNR evaluated at a given time-frequency bin is given by the ratio of the variance of the filtered signal to the variance of the filtered noise
\begin{equation}\label{eq:SNRout}
\begin{aligned}
 \text{SNR}_{\text{out}}=&\frac{\mathbb{E}\left\{|\hat{\mathbf{w}}^H\mathbf{x}|^2 \right\}}{\mathbb{E}\left\{|\hat{\mathbf{w}}^H\mathbf{n}|^2 \right\}}
                        =\frac{\hat{\mathbf{w}}^H \mathbf{R}_{\mathbf{xx}} \hat{\mathbf{w}}}{\hat{\mathbf{w}}^H \mathbf{R}_{\mathbf{nn}} \hat{\mathbf{w}}}\\
                        =& P_s \mathbf{a}^H\mathbf{R}_{\mathbf{nn}}^{-1}\mathbf{a}.
 \end{aligned}
\end{equation}

\subsection{Sensor selection model}\label{sec:ss_model}
The task of sensor selection is to determine the best subset of sensors to activate in order to minimize an objective function, subject to some constraints, e.g., the number of activated sensors or output noise power. We introduce a selection vector
\begin{equation}
 \mathbf{p}=[p_1,p_2,...,p_M]^T \in\{0,1\}^M,
\end{equation}
where $p_i=1$ indicates that the $i$-th sensor is selected. Let $K=\|\mathbf{p}\|_0$ represent the number of selected sensors with the $\ell_0$-(quasi) norm referring to the number of non-zero entries in $\mathbf{p}$. Using a sensor selection matrix $\mathbf{\Phi_p}$, the selected microphone measurements can be compactly expressed as
\begin{equation}
 \mathbf{y}_{\mathbf{p}}=\mathbf{\Phi}_{\mathbf{p}}\mathbf{y}=\mathbf{\Phi}_{\mathbf{p}}\mathbf{x}+\mathbf{\Phi}_{\mathbf{p}}\mathbf{n},
\end{equation}
where $\mathbf{y}_{\mathbf{p}}\in \mathbb{C}^K$ is the vector containing the measurements from the selected sensors. Let $\text{diag}(\mathbf{p})$ be a diagonal matrix whose diagonal entries are given by $\mathbf{p}$, such that $\mathbf{\Phi}_{\mathbf{p}}\in \{0,1\}^{K\times M}$ is a submatrix of $\text{diag}(\mathbf{p})$ after all-zero rows (corresponding to the unselected sensors) have been removed. As a result, we can easily get the following relationships
\begin{equation}
\begin{aligned}
  \mathbf{\Phi}_{\mathbf{p}}\mathbf{\Phi}_{\mathbf{p}}^T  = \mathbf{I}_K, \ \
  \mathbf{\Phi}_{\mathbf{p}}^T\mathbf{\Phi}_{\mathbf{p}}  = \text{diag}(\mathbf{p}).
\end{aligned}
\end{equation}
Therefore, applying the selection model to the classical MVDR beamformer in Sec.~\ref{sec:mvdr}, the best linear unbiased estimator for a subset of $K$ microphones determined by $\mathbf{p}$ will be
\begin{equation}\label{eq:MVDR-ss}
  \hat{\mathbf{w}}_{\mathbf{p}}=\frac{\mathbf{R}_{\mathbf{nn,p}}^{-1}\mathbf{a}_{\mathbf{p}}}{\mathbf{a}_{\mathbf{p}}^H\mathbf{R}_{\mathbf{nn,p}}^{-1}\mathbf{a}_{\mathbf{p}}} ,
\end{equation}
where $\mathbf{a}_{\mathbf{p}} =\mathbf{\Phi}_{\mathbf{p}}\mathbf{a}$ is the ATF of selected microphones, and
$ \mathbf{R}_{\mathbf{nn,p}} = \mathbf{\Phi}_{\mathbf{p}}\mathbf{R}_{\mathbf{nn}}\mathbf{\Phi}_{\mathbf{p}}^T$
represents the noise correlation matrix of the selected sensors after the rows and columns of $\mathbf{R}_{\mathbf{nn}}$ corresponding to the unselected sensors have been removed, i.e., $ \mathbf{R}_{\mathbf{nn,p}}$ is a submatrix of $\mathbf{R}_{\mathbf{nn}}$.

\section{Problem formulation}\label{sec:problem_formulation}
This work focuses on selecting the most informative subset of microphones for MVDR beamforming based noise reduction. The problem is formulated from the viewpoint of minimizing transmission cost subject to a constraint on the output performance.  In particular, we express the filtering performance in terms of the output noise power, which is under the MVDR beamformer equivalent to the output SNR. However, notice that this  can easily be replaced by other performance measures expressing the desired quality or intelligibility.

Let $\mathbf{c} = [c_1, c_2,...,c_M]^T \in \mathbb{R}^M$ denote the pairwise transmission cost between each microphone and the FC. In general, the power consumption for wireless transmission can be modeled as [26]
\begin{equation} \label{eq:cost_model}
 c_i = c(d_i) + c_i^{(0)},\ \forall i,
\end{equation}
where $c(d_i)$ represents the power consumption depending on the distance $d_i$ from the node with the $i$-th microphone to the FC, and $ c_i^{(0)}$ is a constant depending on the power consumption of the $i$-th microphone itself.
Based on the power model of wireless transmission in (\ref{eq:cost_model}), our initial problem can be formulated as
\begin{equation}\tag{P1}
\begin{aligned}
  \min_{\mathbf{w_p},\mathbf{p}\in \{0,1 \}^M} \ \  &\|\text{diag}(\mathbf{p})\mathbf{c} \|_1 \\
  {\rm s. t.} \ \ \ \ &\mathbf{w}^H_{\mathbf{p}}\mathbf{R_{nn,p}}\mathbf{w_p} \leq \frac{\beta}{\alpha},\\
  &\mathbf{w}^H_{\mathbf{p}}\mathbf{a_p}=1,
\end{aligned}
\end{equation}
where $||\cdot||_1$ denotes the $\ell_1$-norm, $\beta$ denotes the output noise power after beamforming when all sensors are used (i.e., $\mathbf{p}=\mathbf{1}_M$), and $\alpha \in(0,1] $ is an adaptive factor to control the output noise power compared to $\beta$. In (P1), the $\ell_1$-norm is used to represent the total transmission costs of the network, i.e., between all the selected sensors and the FC, and it equals the inner-product $\mathbf{c}^T \mathbf{p}$ since both $\bf{p}$ and $\bf{c}$ are non-negative. Also, notice that (P1) is a general case for spatial filtering based noise reduction problems, e.g., using MVDR beamformers or linear constrained minimum variance (LCMV) beamformers [27]. In the next section, we will show how the optimization problem in (P1) can be solved using some of the properties of the MVDR beamformer.

\section{Model-driven sensor selection}\label{sec:model-selection}
In this section, we propose two slightly different ways to solve the optimization problem in (P1), firstly based on the correlation matrix $\mathbf{R_{xx}}$ and secondly based on knowledge of the steering vector $\mathbf{a}$, respectively. Both these solvers rely on the knowledge of the correlation matrices of the complete network, so that they belong to the model-driven schemes.

Considering the MVDR beamformer in (\ref{eq:MVDR-ss}), the output noise power using the selected sensors is given by
\begin{equation}\label{eq:out_npower_mvdr}
\begin{aligned}
   \hat{\mathbf{w}}^H_{\mathbf{p}}\mathbf{R}_{\mathbf{nn,p}} \hat{\mathbf{w}}_{\mathbf{p}}
   =\left(\mathbf{a}_{\mathbf{p}}^H\mathbf{R}_{\mathbf{nn,p}}^{-1}\mathbf{a}_{\mathbf{p}} \right)^{-1},
\end{aligned}
\end{equation}
where the constraint $\mathbf{w}^H_{\mathbf{p}}\mathbf{a_p}=1$ in (P1) is implicit.
Based on the fact that the MVDR beamformer keeps the speech components undistorted and suppresses the noise components, the variance of the filtered speech components can be shown to equal
\begin{equation}
  \hat{\mathbf{w}}^H_{\mathbf{p}}\mathbf{R}_{\mathbf{xx,p}} \hat{\mathbf{w}}_{\mathbf{p}}=P_s,
\end{equation}
where $\mathbf{R}_{\mathbf{xx,p}}$ denotes the submatrix of $\mathbf{R_{xx}}$ corresponding to the selected sensors. Hence, following (\ref{eq:SNRout}) the output SNR using the selected sensors is given by
\begin{equation}\label{eq:SNRp1}
\begin{aligned}
 \text{SNR}_{\text{out},\mathbf{p}} & = \frac{\hat{\mathbf{w}}^H_{\mathbf{p}}\mathbf{R}_{\mathbf{xx,p}} \hat{\mathbf{w}}_{\mathbf{p}}}{ \hat{\mathbf{w}}^H_{\mathbf{p}}\mathbf{R}_{\mathbf{nn,p}} \hat{\mathbf{w}}_{\mathbf{p}}}\\
                         &= P_s \mathbf{a}_{\mathbf{p}}^H\mathbf{R}_{\mathbf{nn,p}}^{-1}\mathbf{a}_{\mathbf{p}}\\
                         &= P_s \mathbf{a}^H\mathbf{\Phi}_{\mathbf{p}}^T\mathbf{R}^{-1}_{\mathbf{nn,p}}\mathbf{\Phi}_{\mathbf{p}}\mathbf{a}.
\end{aligned}
\end{equation}
As a result, the original optimization problem in (P1) can equivalently be rewritten as
\begin{equation}\tag{P2}
\begin{aligned}
  & \min_{\mathbf{p}\in \{0,1 \}^M} \ \  \|\text{diag}(\mathbf{p})\mathbf{c} \|_1 \\
  & {\rm s. t.} \ \ P_s\mathbf{a}_{\mathbf{p}}^H\mathbf{R}_{\mathbf{nn,p}}^{-1}\mathbf{a}_{\mathbf{p}} \ge {\alpha \cdot \text{SNR}},
\end{aligned}
\end{equation}
where $\text{SNR}=\frac{P_s}{\beta}$ represents the output SNR when all the sensors are used. Both (P1) and (P2) are non-convex because of the Boolean variable $\mathbf{p}$, but also due to the non-linearity of the constraint in $\mathbf{p}$. In what follows, we will present solvers by linearizing (P2) and reformulating it using convex relaxation. Note that (P1) and (P2) are built from different perspectives (i.e., constraining the output noise power and SNR, respectively), but in the context of the MVDR beamforming, they are equivalent.
\subsection{Convex relaxation using $\mathbf{R_{xx}}$}
From the output SNR in (\ref{eq:SNRp1}), the selection variable $\mathbf{p}$ appears at three places, that are: $\mathbf{\Phi}_{\mathbf{p}}^T$, $\mathbf{R}^{-1}_{\mathbf{nn,p}}$ and $\mathbf{\Phi}_{\mathbf{p}}$. We combine these together as one new matrix $\mathbf{Q}=\mathbf{\Phi}_{\mathbf{p}}^T\mathbf{R}^{-1}_{\mathbf{nn,p}}\mathbf{\Phi}_{\mathbf{p}}$. To simplify calculations, in what follows, we will rearrange $\mathbf{Q}$ such that $\mathbf{p}$ occurs only at one place.
Let us first consider a decomposition of the noise covariance matrix [14][28]
\begin{equation}\label{eq:decompRnn}
  \mathbf{R_{nn}} = \lambda \mathbf{I}_M + \mathbf{G},
\end{equation}
where $\lambda$ is a positive scalar and $\mathbf{G}$ is a positive definite matrix (if $\lambda$ is smaller than the smallest eigenvalue of $\mathbf{R_{nn}}$, this decomposition can be easily found). The reason for choosing such a $\lambda$ is to make $\mathbf{G}^{-1}+\lambda^{-1}\text{diag}(\mathbf{p})$ positive definite, which will be seen  after (\ref{eq:constraint4}).
Using (\ref{eq:decompRnn}), we have
\begin{equation}
\begin{aligned}
  \mathbf{R}_{\mathbf{nn,p}} = \mathbf{\Phi}_{\mathbf{p}}\left(\lambda\mathbf{I}_M + \mathbf{G} \right) \mathbf{\Phi}_{\mathbf{p}}^T
                             = \lambda \mathbf{I}_K + \mathbf{\Phi}_{\mathbf{p}} \mathbf{G} \mathbf{\Phi}_{\mathbf{p}}^T,
\end{aligned}
\end{equation}
and $\mathbf{Q}$ can be reformulated as
\begin{equation}\label{eq:defineA}
\begin{aligned}
  \mathbf{Q} = \mathbf{\Phi}_{\mathbf{p}}^T\left(\lambda \mathbf{I}_K + \mathbf{\Phi}_{\mathbf{p}} \mathbf{G} \mathbf{\Phi}_{\mathbf{p}}^T \right)^{-1} \mathbf{\Phi}_{\mathbf{p}}.
\end{aligned}
\end{equation}
Using the matrix inversion lemma~[29, p.18]
\begin{equation}\nonumber
\mathbf{C} \left(\mathbf{B}^{-1}+\mathbf{C}^T\mathbf{A}^{-1}\mathbf{C} \right)^{-1}\mathbf{C}^T=\mathbf{A}-\mathbf{A}\left(\mathbf{A}+\mathbf{CBC}^T\right)^{-1}\mathbf{A},
\end{equation}
we can simplify $\mathbf{Q}$ in (\ref{eq:defineA}) as
\begin{equation}\label{eq:constructA}
\begin{aligned}
 \mathbf{Q} = \mathbf{G}^{-1}-\mathbf{G}^{-1}\left(\mathbf{G}^{-1}+\lambda^{-1}\text{diag}(\mathbf{p}) \right)^{-1}\mathbf{G}^{-1}.
\end{aligned}
\end{equation}
Note that (\ref{eq:constructA}) is still non-linear in $\mathbf{p}$ due to the inversion operation, but $\bf{p}$ appears now only at one place.
Based on $\mathbf{Q}$, the output SNR with sensor selection as in (\ref{eq:SNRp1}) can be calculated as
\begin{equation}\label{eq:SNRp2}
\begin{aligned}
 \text{SNR}_{\text{out},\mathbf{p}} &\overset{\text{(1)}}{=} \text{trace}\left(P_s\mathbf{a}^H\mathbf{\Phi}_{\mathbf{p}}^T\mathbf{R}_{\mathbf{nn,p}}^{-1}\mathbf{\Phi}_{\mathbf{p}}\mathbf{a}\right)\\
 &\overset{\text{(2)}}{=} \text{trace}\left(\mathbf{QR_{xx}}\right)\\
 & \overset{\text{(3)}}{=} \text{trace}\left(\mathbf{R}_{\mathbf{xx}}^{\frac{H}{2}} \mathbf{Q} \mathbf{R}_{\mathbf{xx}}^{\frac{1}{2}}\right),
\end{aligned}
\end{equation}
where the $\text{trace}(\cdot)$ operator computes the trace of a matrix, and $\mathbf{R}_{\mathbf{xx}}^{\frac{1}{2}}$ represents the principle square root of $\mathbf{R_{xx}}$. The second and third equality in (\ref{eq:SNRp2}) is based on trace property, which is employed to make the linear matrix inequality (LMI) in (\ref{eq:constraint5}) symmetric.  Here, we utilize the trace operation to express the output SNR as a function of $\mathbf{R_{xx}}$. The latter can be estimated using the recorded audio in practice, e.g., during the training phase, or using the correlation matrices $\mathbf{R_{yy}}$ and $\mathbf{R_{nn}}$ without the need to explicitly know the steering vector $\mathbf{a}$ or $\mathbf{a_p}$.

Secondly, in what follows we will linearize the SNR constraint in (P2). To do this, we introduce a new matrix $\mathbf{Z}$ to equivalently rewrite the constraint in (P2) as
\begin{gather}
  \text{trace}\left(\mathbf{Z}- \frac{\alpha P_s}{M \beta} \mathbf{I}_M \right) \ge 0 \label{eq:constraint1},\\
  \mathbf{R}_{\mathbf{xx}}^{\frac{H}{2}} \mathbf{Q} \mathbf{R}_{\mathbf{xx}}^{\frac{1}{2}} = \mathbf{Z} \label{eq:constraint2},
\end{gather}
where the equality constraint in (\ref{eq:constraint2}) is non-linear in $\mathbf{p}$. For linearization, we relax it to an inequality constraint
\begin{equation}\label{eq:constraint3}
 \mathbf{R}_{\mathbf{xx}}^{\frac{H}{2}} \mathbf{Q} \mathbf{R}_{\mathbf{xx}}^{\frac{1}{2}} \succeq \mathbf{Z}.
\end{equation}
Note that (\ref{eq:constraint1}) and (\ref{eq:constraint3}) are sufficient conditions for obtaining the original constraint in (P2), this is why we utilize $\succeq$ for convex relaxation. Substituting (\ref{eq:constructA}) in (\ref{eq:constraint3}), we get
\begin{equation}\label{eq:constraint4}
\begin{aligned}
\mathbf{R}_{\mathbf{xx}}^{\frac{H}{2}}\mathbf{G}^{-1} &\mathbf{R}_{\mathbf{xx}}^{\frac{1}{2}}-\mathbf{Z}\succeq \\ &\mathbf{R}_{\mathbf{xx}}^{\frac{H}{2}}\mathbf{G}^{-1}\left[\mathbf{G}^{-1}+\lambda^{-1}\text{diag}\left(\mathbf{p}\right)\right]^{-1}\mathbf{G}^{-1}\mathbf{R}_{\mathbf{xx}}^{\frac{1}{2}}.
\end{aligned}
\end{equation}
Due to the positivity of $\lambda$, the positive definiteness of $\mathbf{G}$ and the Boolean vector $\mathbf{p}$, the matrix $\mathbf{G}^{-1}+\lambda^{-1}\text{diag}(\mathbf{p})$ is positive definite, and this is why we chose in (\ref{eq:decompRnn}) a positive scalar $\lambda$ and a positive definite matrix $\mathbf{G}$ to decompose the matrix $\mathbf{R_{nn}}$. Using the Schur complement~[30, p.650], we obtain a symmetric LMI of size $2M$ from (\ref{eq:constraint4}) as
\begin{equation}\label{eq:constraint5}
        \begin{bmatrix}\mathbf{G}^{-1}+\lambda^{-1}\text{diag}(\mathbf{p})  & \mathbf{G}^{-1}\mathbf{R}_{\mathbf{xx}}^{\frac{1}{2}}\\ \mathbf{R}_{\mathbf{xx}}^{\frac{H}{2}}\mathbf{G}^{-1}   & \mathbf{R}_{\mathbf{xx}}^{\frac{H}{2}}\mathbf{G}^{-1}\mathbf{R}_{\mathbf{xx}}^{\frac{1}{2}}-\mathbf{Z} \end{bmatrix} \succeq \mathbf{0}_{2M} ,
\end{equation}
 which is linear in $\mathbf{p}$. Furthermore, the Boolean variable $\mathbf{p}$ can be relaxed using continuous variables $\mathbf{p}\in[0,1]^M$ or semidefinite relaxation [31]. In this work, we utilize the former way.
Accordingly, (P2) can be expressed in the following form:
\begin{equation}\label{eq:optimization2}
\begin{aligned}
  \min_{\mathbf{p}, \mathbf{Z}} \ \ &\|\text{diag}(\mathbf{p})\mathbf{c} \|_1 \\
  {\rm s. t.} \ \  &\text{trace}\left(\mathbf{Z}- \frac{\alpha P_s}{M \beta} \mathbf{I}_M \right) \ge 0, \\
                       & \begin{bmatrix}\mathbf{G}^{-1}+ \lambda^{-1}\text{diag}(\mathbf{p})  & \mathbf{G}^{-1}\mathbf{R}_{\mathbf{xx}}^{\frac{1}{2}}\\ \mathbf{R}_{\mathbf{xx}}^{\frac{H}{2}}\mathbf{G}^{-1}   & \mathbf{R}_{\mathbf{xx}}^{\frac{H}{2}}\mathbf{G}^{-1}\mathbf{R}_{\mathbf{xx}}^{\frac{1}{2}}-\mathbf{Z} \end{bmatrix} \succeq \mathbf{0}_{2M},\\
                        & 0\leq p_i \leq 1, \ \  i=1,2,\cdots,M.
\end{aligned}
\end{equation}
The relaxed optimization problem in (\ref{eq:optimization2}) is a semidefinite programming problem~[30, p.128] and can be solved efficiently in polynomial time using interior-point methods or solvers,  like CVX~[32] or SeDuMi~[33]. The computational complexity for solving (\ref{eq:optimization2}) is of the order of $\mathcal{O}(M^3)$. The approximate Boolean selection variables $p_i$ can be obtained by randomized rounding using the solution of (\ref{eq:optimization2})~[13].
Notice that the solver in (\ref{eq:optimization2}) depends on $\mathbf{R}_{\mathbf{xx}}$. In a practical scenario, this is unknown, but can be estimated based on estimates of the correlation matrices $\mathbf{R}_{\mathbf{yy}}$ and $\mathbf{R}_{\mathbf{nn}}$ as shown in (\ref{eq:Ryy=Rxx+Rnn}). $\mathbf{R}_{\mathbf{yy}}$ can be estimated from the data itself, and $\mathbf{R}_{\mathbf{nn}}$ can be estimated using a voice activity detector or noise correlation matrix estimator, see e.g.,[34].

\subsection{Solver based on the steering vector $\mathbf{a}$}
Assuming that the relative locations of the sensors and sources are known, the steering vectors $\mathbf{a}$ (in free field) are available. Given $\mathbf{a}$, the output SNR in (\ref{eq:SNRp2}) can be represented as
 \begin{equation}\label{SNRwith_a}
   \text{SNR}_{\text{out},\mathbf{p}}=P_s\mathbf{a}^H\mathbf{Qa}.
 \end{equation}
Therefore, using (\ref{eq:constructA}), the original constraint in (P2) can be rewritten as
\begin{equation}\label{eq:constraint6}\nonumber
\begin{aligned}
 \mathbf{a}^H\mathbf{G}^{-1}\mathbf{a}-\mathbf{a}^H\mathbf{G}^{-1}\left(\mathbf{G}^{-1}+\lambda^{-1}\text{diag}(\mathbf{p}) \right)^{-1}\mathbf{G}^{-1}\mathbf{a} \ge \frac{\alpha}{ \beta},
\end{aligned}
\end{equation}
or, reorganized as
\begin{equation}\label{eq:constraint7}
\begin{aligned}
 \mathbf{a}^H\mathbf{G}^{-1}\mathbf{a}-\frac{\alpha}{ \beta} \ge \mathbf{a}^H\mathbf{G}^{-1}\left(\mathbf{G}^{-1}+\lambda^{-1}\text{diag}(\mathbf{p}) \right)^{-1}\mathbf{G}^{-1}\mathbf{a}.
\end{aligned}
\end{equation}
Using the Schur complement, (\ref{eq:constraint7}) can be reformulated as a symmetric LMI of size $M+1$
\begin{equation}\label{eq:constraint8}
       \begin{bmatrix}\mathbf{G}^{-1}+\lambda^{-1}\text{diag}(\mathbf{p})  & \mathbf{G}^{-1}\mathbf{a}\\ \mathbf{a}^H\mathbf{G}^{-1}   & \mathbf{a}^H\mathbf{G}^{-1}\mathbf{a}- \frac{\alpha}{ \beta}\end{bmatrix} \succeq \mathbf{0}_{M+1}.
\end{equation}
Accordingly, the optimization problem in (P2) is expressed as
\begin{equation}\label{eq:optimization3}
\begin{aligned}
  \min_{\mathbf{p} }\ \  &\|\text{diag}(\mathbf{p})\mathbf{c} \|_1 \\
  {\rm s. t.} \ \ & \begin{bmatrix}\mathbf{G}^{-1}+\lambda^{-1}\text{diag}(\mathbf{p})  & \mathbf{G}^{-1}\mathbf{a}\\ \mathbf{a}^H\mathbf{G}^{-1}   & \mathbf{a}^H\mathbf{G}^{-1}\mathbf{a}- \frac{\alpha}{ \beta}\end{bmatrix} \succeq \mathbf{0}_{M+1}\\
  & 0\leq p_i \leq 1, \ \  i=1,2,\cdots,M,
\end{aligned}
\end{equation}
where the Boolean variables $\mathbf{p}$ have already been relaxed using the continuous surrogates $\mathbf{p}\in[0,1]^M$, and (\ref{eq:optimization3}) has a standard semidefinite programming form, which can also be solved by the aforementioned tools. Notice that this solver depends on knowledge on $\bf{a}$. To estimate (the direct path of)  $\bf{a}$ one can use
a source localization algorithm, e.g.,~[35]-[37], in combination with the sensor locations, or use the generalized eigenvalue decomposition of the matrices $\mathbf{R_{nn}}$
and $\mathbf{R_{yy}}$~[38][39].
\begin{remark}
  The differences between (\ref{eq:optimization2}) and (\ref{eq:optimization3}) are threefold: 1) (\ref{eq:optimization3}) preserves the constraint on the output SNR (or noise power), yet (\ref{eq:optimization2}) relaxes it in a convex way by introducing an auxiliary variable $\mathbf{Z}$; 2) Observing the LMIs in (\ref{eq:optimization2}) and (\ref{eq:optimization3}), they differ in dimensions (i.e., $2M$ and $M+1$, respectively), so (\ref{eq:optimization3}) is computationally much more efficient; 3) The solver in (\ref{eq:optimization2}) requires to estimate the speech correlation matrix $\mathbf{R_{xx}}$ and the PSD $P_s$ of the target source, while (\ref{eq:optimization3}) requires the steering vector $\mathbf{a}$.
\end{remark}

\begin{remark}
For a special case, when the noise is spatially uncorrelated with covariance matrix
\[
\mathbf{R}_\mathbf{nn}={\rm diag}\left(\sigma_1^{2}, \sigma_2^{2},\cdots,\sigma_M^{2}\right),
\]
the optimization problem (P2) can be simplified to the following Boolean linear programming problem
 \begin{equation}
 \label{eq:uncorrelated}
 \begin{aligned}
  &{\displaystyle \min_{\mathbf{p} }}   \quad \|{\rm diag}(\mathbf{p})\mathbf{c} \|_1\\
  &{\rm s. t.}     \quad \mathbf{a}^H \mathbf{R}^{-1}_\mathbf{nn}{\rm diag}(\mathbf{p})\mathbf{a} \geq \frac{\alpha }{\beta}.
 \end{aligned}
 \end{equation}
Although the above optimization problem is nonconvex in $\mathbf{p} \in \{0,1\}^M$, it admits a simple non-iterative solution based on rank ordering. More specifically, the optimal solution to \eqref{eq:uncorrelated} is given by setting the entries of ${\bf p}$ corresponding to the indices
\[
\min \  \left\{i \in \{1,2,\cdots,M\} \vert \frac{c_{[1]}}{v_{[1]}}  + \cdots +\frac{c_{[i]}}{v_{[i]}} \geq \frac{\alpha}{\beta}\right\}
\]
to 1, and the remaining entries of ${\bf p}$ to 0, where $v_{[1]}, \cdots,v_{[M]}$ and $c_{[1]}, \cdots,c_{[M]}$ are numbers of $v_{1}, v_{2}, \cdots,v_{M}$ and $c_{1}, c_{2}, \cdots,c_{M}$, respectively, sorted in ascending order with $v_i =c_i \sigma_i^{2} /|a_i|^2$ and $a_i$ being the $i$-th entry of ${\bf a}$.
\end{remark}

\section{Greedy sensor selection}\label{sec:greedy}
In Sec.~\ref{sec:model-selection}, the sensor selection problem was solved using statistical information from the complete network, i.e., $\mathbf{R_{xx}}$ and $\mathbf{R_{nn}}$. In practice, this information is unknown  and needs to be estimated from all the sensors' measurements. Hence, we call this a model-driven approach as the complete $\mathbf{R_{xx}}$ and $\mathbf{R_{nn}}$ are required as well as the transmission power from the microphones to the FC. In a practical scenario, it is undesired to estimate the statistics of the complete network up front, as this would imply a lot of data transmission for sensor nodes that might never be selected in the end as most sensors are non-informative.
Moreover, in practice, the position of the FC or microphones might be changing as well. For this reason we need a selection mechanism that does not rely on knowledge of the statistics and microphone-FC distances  of the complete network.
Instead, we could access the measurements of neighboring sensors (close to the FC or already selected sensors). In this section, we present a greedy approach for the sensor selection based noise reduction problem, which does not require to estimate the global statistics. Therefore, the greedy algorithm can be performed online, and it belongs to the data-driven category. In Sec.~\ref{sec:exp}, we will experimentally show that the data-driven and model-driven approach will converge to each other.

Let $r_i$ denote the spatial position of the $i$-th microphone, $\mathcal{S}_1$ a candidate set of microphones and $\mathcal{S}_2$ the selected set, respectively. The proposed greedy algorithm is summarized in Algorithm~\ref{alg:improved_greedy}. Given an arbitrary initial spatial point $z_0$ and a transmission range $R_0$\footnote{$R_0$ can be defined as the wireless transmission range $\sqrt{\log (2M)/M}$ in a random geometric graph to guarantee that the network is connected with high probability~[40].}, we can initialize the candidate set $\mathcal{S}_1$ of sensors, i.e., the $R_0$-closest sensors to $z_0$. For the candidate set $\mathcal{S}_1$, we estimate the noise correlation matrix $\mathbf{R_{\mathbf{nn},\mathcal{S}_1}}$ and decompose it following (\ref{eq:decompRnn}), and then solve the optimization problem in (\ref{eq:optimization2}) or (\ref{eq:optimization3}). For instance, for $\mathcal{S}_1$ (\ref{eq:optimization3}) can be reformulated as
 \begin{equation}\label{eq:greedy_optimize}
 \begin{aligned}
       & \min_{\mathbf{p}\in[0,1]^{K_1}}\ \ \|\text{diag}(\mathbf{p}_{\mathcal{S}_1})\mathbf{c}_{\mathcal{S}_1} \|_1 \\
     &  {\rm s. t.} \ \  \begin{bmatrix}\mathbf{G}^{-1}_{\mathcal{S}_1}+\lambda^{-1}_{\mathcal{S}_1}\text{diag}(\mathbf{p}_{\mathcal{S}_1})  & \mathbf{G}^{-1}_{\mathcal{S}_1}\mathbf{a}_{\mathcal{S}_1}\\ \mathbf{a}^H_{\mathcal{S}_1}\mathbf{G}^{-1}_{\mathcal{S}_1}   & \mathbf{a}^H_{\mathcal{S}_1}\mathbf{G}^{-1}_{\mathcal{S}_1}\mathbf{a}_{\mathcal{S}_1}- \frac{\alpha}{ \beta_{\mathcal{S}_1}}\end{bmatrix} \succeq \mathbf{0}_{K_1+1}\\
      &\hskip 3em 0 \leq p_i \leq 1,\ \forall i \in \mathcal{S}_1,
 \end{aligned}
 \end{equation}
where $\beta_{\mathcal{S}_1}$ represents the output noise power of the classical MVDR beamformer using the microphones in the candidate set ${\mathcal{S}_1}$, which is termed as the local constraint. Notice that the adaptive factor $\alpha$ is the same as that in the model-driven scheme.
If $\alpha \leq 1$, (\ref{eq:greedy_optimize}) will always have a feasible solution within $\mathcal{S}_1$, the feasible set will be taken and used to define a new set $\mathcal{S}_2$ with $|{\mathcal{S}_2}| \leq |{\mathcal{S}_1}|$. Then, based on the set $\mathcal{S}_2$, a new set $\mathcal{S}_1$ is formed based on the $R_0$-closest sensors with respect to the sensors included in the set $\mathcal{S}_2$\footnote{$R_0$-closest sensors with respect to the set $\mathcal{S}_2$ include all the sensors that are $R_0$-closest to any individual sensor in $\mathcal{S}_2$.}. These operations are continued until $\mathcal{S}_1$ or $\mathcal{S}_2$ does not change (i.e., until convergence has been achieved). The finally selected set $\mathcal{S}_2$ will always be smaller than the selected set for the model-driven approach from Sec.~\ref{sec:model-selection} when using the same $\alpha$. This is due to the fact that the output noise power $\beta_{\mathcal{S}_1}$ in the constraint of the greedy approach is based on the set $\mathcal{S}_1$ that is always smaller or equal to the initial set as used  by the model-driven approach in (\ref{eq:optimization3}) (which simply contains all sensors). As a result, $\beta/\alpha$ will always be smaller than $\beta_{\mathcal{S}_1}/\alpha$. In summary,  $\beta/\alpha  < \beta_{\mathcal{S}_1}/\alpha$. The performance of the greedy approach (after convergence) will therefore always be somewhat worse than the model-based approach, as the constraint is less tight. This can either be solved by choosing a different  (larger) $\alpha$ for the greedy approach, or, by switching from the constraint  $\beta_{\mathcal{S}_1}/\alpha$ to the constraint $\beta/\alpha$ after convergence.
As an alternative, we could have used the constraint $\beta/\alpha $ within the greedy approach of (\ref{eq:greedy_optimize}) right from the beginning. However, in that case,  in the few several iterations (\ref{eq:greedy_optimize}) would have no feasible solution as an insufficient amount of measurements are available to satisfy the constraint on the output noise power. As a consequence of an infeasible solution, the selected set $\mathcal{S}_2$ will keep all sensors from $\mathcal{S}_1$, of which many are actually uninformative.

To make the performance of the proposed greedy algorithm converge to the model-driven approach,  we switch from $\beta_{\mathcal{S}_1}$ (local constraint) to $\beta$ (global constraint)  after the above iterative procedure
converges (i.e., the constraint $\beta_{\mathcal{S}_1}/\alpha$ for solving (\ref{eq:greedy_optimize}) has been satisfied).
Finally, the proposed greedy algorithm will converge to the model-driven method based on the global constraint. To conclude, the greedy algorithm includes two steps: using a locally defined constraint ($\beta_{\mathcal{S}_1}/\alpha$) and using a globally defined constraint ($\beta/\alpha$), as summarized in Algorithm 1. Note that the computational complexity of each iteration is of the order of $\mathcal{O}(|\mathcal{S}_1|^3)$, and the number of iterations depends on $z_0$ and $R_0$.
\begin{algorithm}[!t]
\caption{Greedy Sensor Selection}
\label{alg:improved_greedy}
\textbf{Initialization:} \\
   \hskip 3em Initial point: $z_0$\\
   \hskip 3em Transmission range: $R_0$ \\
   \hskip 3em Selected set: $\mathcal{S}_2=\varnothing$\\
   \hskip 3em Candidate set: $\mathcal{S}_1 =\{i| \|r_i -z_0 \|_2\leq R_0,\ \forall i$ \};\\
  Cardinality of the active set: $K_1=|\mathcal{S}_1|;$ \\
   Decomposing: $\mathbf{R_{\mathbf{nn},\mathcal{S}_1}} = \lambda_{\mathcal{S}_1} \mathbf{I}_{K_1} + \mathbf{G}_{\mathcal{S}_1} \in \mathbb{C}^{{K_1}\times {K_1}}$;\\
   \textbf{Solving (\ref{eq:greedy_optimize})} using the local constraint $\beta_{\mathcal{S}_1}$; \\
   \textbf{Update:}\\
     \hskip 3em  $\mathcal{S}_2=\{i|p_i=1, \ \forall i\in \mathcal{S}_1$\}; \\
     \hskip 3em  $\mathcal{S}_1 = \mathcal{S}_2 \cup \{i|\|r_i - r_{\mathcal{S}_2} \|_2\leq R_0 ,\ \forall i\}$;\\
     \hskip 3em  \textbf{Go to step 6 until converge};\\
   \textbf{Solving (\ref{eq:greedy_optimize})} using global constraint $\beta$; \\
      \textbf{If infeasible, update}\\
     \hskip 3em  $\mathcal{S}_2=\mathcal{S}_1$;\\
     \hskip 3em  $\mathcal{S}_1 = \mathcal{S}_2 \cup \{i|\|r_i - r_{\mathcal{S}_2} \|_2\leq R_0 ,\ \forall i\}$;\\
     \hskip 3em  \textbf{Go to step 13};\\
   \textbf{If feasible, update}\\
     \hskip 3em  $\mathcal{S}_2=\{i|p_i=1, \ \forall i\in \mathcal{S}_1$\}; \\
     \hskip 3em  $\mathcal{S}_1 = \mathcal{S}_2 \cup \{i|\|r_i - \mathcal{S}_2 \|_2\leq R_0 ,\ \forall i\}$;\\
     \hskip 3em  \textbf{Go to step 13 until converge};\\
   \textbf{Return} $\mathcal{S}_2$.\\
\end{algorithm}
\section{Simulations}\label{sec:exp}
In this section, the proposed algorithms are experimentally evaluated. Sec.~\ref{sec:refer_method} introduces three reference methods that we will use for comparison. In Sec.~\ref{sec:exp_setup}, the experimental setup is explained. In Sec.~\ref{sec:exp:model-driven}, the proposed model-driven sensor selection based MVDR beamformer (referred to as MD-MVDR in short) is compared with the reference methods introduced in Sec.~\ref{sec:refer_method}. In Sec.~\ref{sec:exp:greedy-method}, we will analyze the performance of the proposed greedy approach as a data-driven sensor selection, including the convergence behaviour, initialization and the adaptivity of a moving FC. Sec.~\ref{sec:exp:complexity} compares the computational complexity between the model-driven method and the greedy approaches.

\subsection{Reference methods}\label{sec:refer_method}

Apart from the classical MVDR beamforming without sensor selection as introduced in Sec.~\ref{sec:mvdr}, the proposed approaches will also be compared with a weighted sparse MVDR beamformer~[41]-[43], a radius-based MVDR beamformer and a utility-based greedy method~[19][20].

\subsubsection{Weighted sparse MVDR beamformer}\label{sec:sparseMVDR}
A naive alternative to sensor selection for spatial filtering is to enforce sparsity in the filter coefficients while designing the beamformer.
Due to the physical nature of sound, this approach trades a
small loss in SNR for a large reduction in communication power required to produce a beamformer output by reducing the
active nodes. Some existing works on sparse MVDR beamformers are presented in [41]-[43].
One of our reference methods is therefore a sparse MVDR beamformer. However in order to make the comparison with the sparse  MVDR beamformer fair, we use a weighting by the transmission power. Using the model of transmission costs from (\ref{eq:cost_model}), the weighted sparse MVDR beamformer can be formulated as
\begin{equation}\label{eq:optimi_sparse}
\begin{aligned}
  \hat{\mathbf{w}} = & \arg\min_{\mathbf{w}} \  \mathbf{w}^H\mathbf{R_{nn}}\mathbf{w} + \mu \|\mathbf{w}^H \text{diag}(\mathbf{c})\|_0 \\
                     & {\rm s. t.} \ \ \mathbf{w}^H\mathbf{a}=1,
\end{aligned}
\end{equation}
where $\mu$ denotes the regularization parameter to control sparsity, and the $\ell_0$-norm can be relaxed by the $\ell_1$-norm or the concave surrogate based on sum-of-logarithms~[13][44]. When $\mu=0$, it is identical to the classical MVDR beamformer in Sec.~\ref{sec:mvdr}. Note that a larger $\mu$ leads to a sparser $\mathbf{w}$.  The product $\mathbf{w}^H \text{diag}(\mathbf{c})$ indicates the pairwise transmission costs. Weighting the beamforming filter $\mathbf{w}$, the sensors with smaller transmission costs have a dominant contribution to $\mathbf{w}$ compared to sensors with larger transmission costs. From the standpoint of implementation, for each frequency bin, if $|w_i|\ge \varepsilon ,\forall i$, the $i$-th sensor will be selected, otherwise not. Due to this ``inevitable " thresholding, the resulting beamformer is not necessarily MVDR anymore. The threshold $\varepsilon$ is chosen empirically.
\subsubsection{Radius-based MVDR beamformer}

The goal of this article is to minimize the transmission costs while constraining the performance. An easy way to reduce transmission costs is by selecting the sensors close to the FC. The closer a sensor to the FC, the less transmission power is required. Hence, given a radius $\gamma$, we can involve the sensors within the circle centered by the FC for the MVDR beamfomer, which we call radius-based MVDR beamformer. An example is given in Fig.~\ref{fig:example_radius_selection}, where the blue sensors are chosen with $\gamma=6$ m. Obviously, this approach does not take the source or interference information into account, and its performance suffers from $\gamma$ and the network topology.
\subsubsection{Utility based greedy sensor addition}\label{sec:utility_greedy_alg}
In~[20], the most informative subset of microphones is obtained by greedily removing the sensor that has the least contribution to a utility measurement (e.g., SNR gain, output noise power, MSE cost), also called backward selection. This method requires to know the statistics offline and can be considered a model-driven approach. While in~[19], apart from sensor selection based on backward selection, an alternative was proposed  by greedily adding the sensor that has the largest contribution to the utility (forward selection). This can be considered as an online data-driven procedure. In order to compare the proposed greedy algorithm with the state-of-the-art greedy methods, we summarize~[19][20] as the utility based greedy sensor addition shown in Algorithm~\ref{alg:utility_greedy}. In this work, our focus is on the transmission costs. To measure the utility, we therefore take the ratio of the gain  of the output noise power $\boldsymbol{\Delta}$ that is obtained by adding each sensor from $\mathcal{S}_1\backslash \mathcal{S}_2 $ to $\mathcal{S}_2$, to the transmission cost. Here, the sets $\mathcal{S}_1$ and $\mathcal{S}_2$, are respectively, defined the same as for Algorithm 1. The sensor which has the larger ratio between noise reduction and transmission cost would have the larger utility. When the transmission costs for the selected set $\mathcal{S}_2$ exceeds the maximum cost budget $c_T$, the algorithm is terminated. Note that this approach only adds one sensor to the selected set $\mathcal{S}_2$ per iteration, thus it may require many iterations to get an acceptable solution.
\begin{algorithm}[!t]
\caption{Utility based greedy sensor addition}
\label{alg:utility_greedy}
\textbf{Initialization:} same to \textbf{Algorithm 1};\\
\textbf{for} $k=1,2,...,M$ \\
   \hskip 2em Compute the gain of output noise power $\boldsymbol{\Delta}$ by adding each sensor in $\mathcal{S}_1\backslash \mathcal{S}_2 $ to $\mathcal{S}_2$; \\
   \hskip 2em Compute utility vector: $\mathbf{g}=[\frac{\Delta_1}{c_1},\frac{\Delta_2}{c_2},...,\frac{\Delta_{|\mathcal{S}_1\backslash \mathcal{S}_2|}}{c_{\mathcal{S}_1\backslash \mathcal{S}_2}}]^T$;\\
   \hskip 2em $i = \arg\max_i \mathbf{g}$;\\
   \hskip 2em Add sensor: $\mathcal{S}_2 = \mathcal{S}_2 \cup i$;\\
   \hskip 2em Update: $\mathcal{S}_1 = \mathcal{S}_2 \cup \{i|\|r_i - r_{\mathcal{S}_2} \|_2\leq R_0 ,\ \forall i\}$;\\
\textbf{end for until} $c_{\mathcal{S}_2} \ge c_T$ \\
\textbf{Return} ${\mathcal{S}_2}$.
\end{algorithm}

\subsection{Experiment Setup}\label{sec:exp_setup}
Fig.~\ref{fig:example_radius_selection} shows the experimental setup employed in the simulations, where 169 candidate microphones are placed uniformly in a 2D room with dimensions $(12\times 12)$ m. The desired speech source (red solid circle) is located at $(2.4, 9.6)$ m. The FC (black solid square) is placed at $(9, 3)$ m. Two interfering sources (blue stars) are positioned at $(2.4, 2.4)$ m and $(9.6, 9.6)$ m, respectively. The target source signal is a 10 minute long concatenation of speech signals originating from the TIMIT database~[45]. The interferences are stationary Gaussian speech shaped noise sources. All signals are sampled at 16 kHz. We use a square-root Hann window of 20 ms for framing with 50\% overlap. The ATFs are generated using~[46] with reverberation time $T_{60}=200$ ms. The threshold $\varepsilon$ for the sparse MVDR beamformer is set to be $10^{-5}$ empirically, since the coefficients smaller than this threshold are negligible. We also model microphone self noise using zero-mean uncorrelated Gaussian noise with an SNR of 50 dB.

To focus on the concept of sensor selection, we assume that the steering vector $\mathbf{a}$ is known. In practice, this can be estimated using source localization algorithms, e.g.,~[35][36], in combination with the sensor locations, or, by calculating the generalized eigenvalue decomposition of the matrices $\mathbf{R_{nn}}$ and $\mathbf{R_{yy}}$~[38][39].
For the wireless transmission model in (\ref{eq:cost_model}), we consider the simplest wireless transmission case, where the transmission cost between each sensor and the FC is proportional to the square of their Euclidean distance~[47], and we assume that the device dependent cost $c_i^{(0)}=0,\forall i$. In the following simulations, the transmission costs are normalized between 0 and 1 based on the total transmission costs between all the microphones and the FC.

\subsection{Evaluation of the model-driven approach}\label{sec:exp:model-driven}
\begin{figure}[!t]
  \centering
  \vspace{-0.5cm}
  \includegraphics[width=0.45\textwidth]{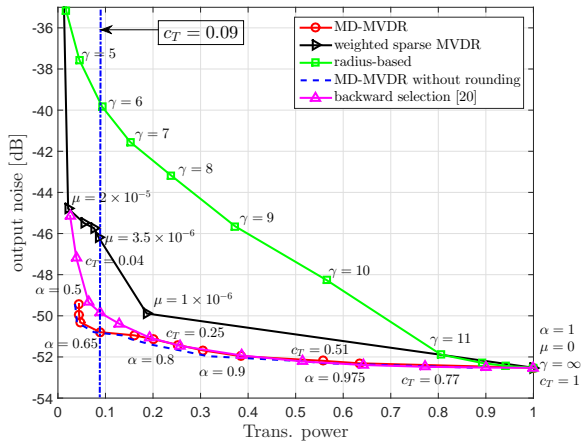}\\
  \vspace{-0.3cm}
  \caption{Output noise power in terms of transmission cost for different choices of $\alpha,\mu, \gamma, c_T$.}
  \label{fig:alpha_mu_radius}
  \vspace{-0.5cm}
\end{figure}

\begin{figure*}[!htb]
\centering
  \begin{subfigure}[Radius-based MVDR ($\gamma=6$)]
      {\includegraphics[width=0.35\textwidth, height=0.3\textwidth]{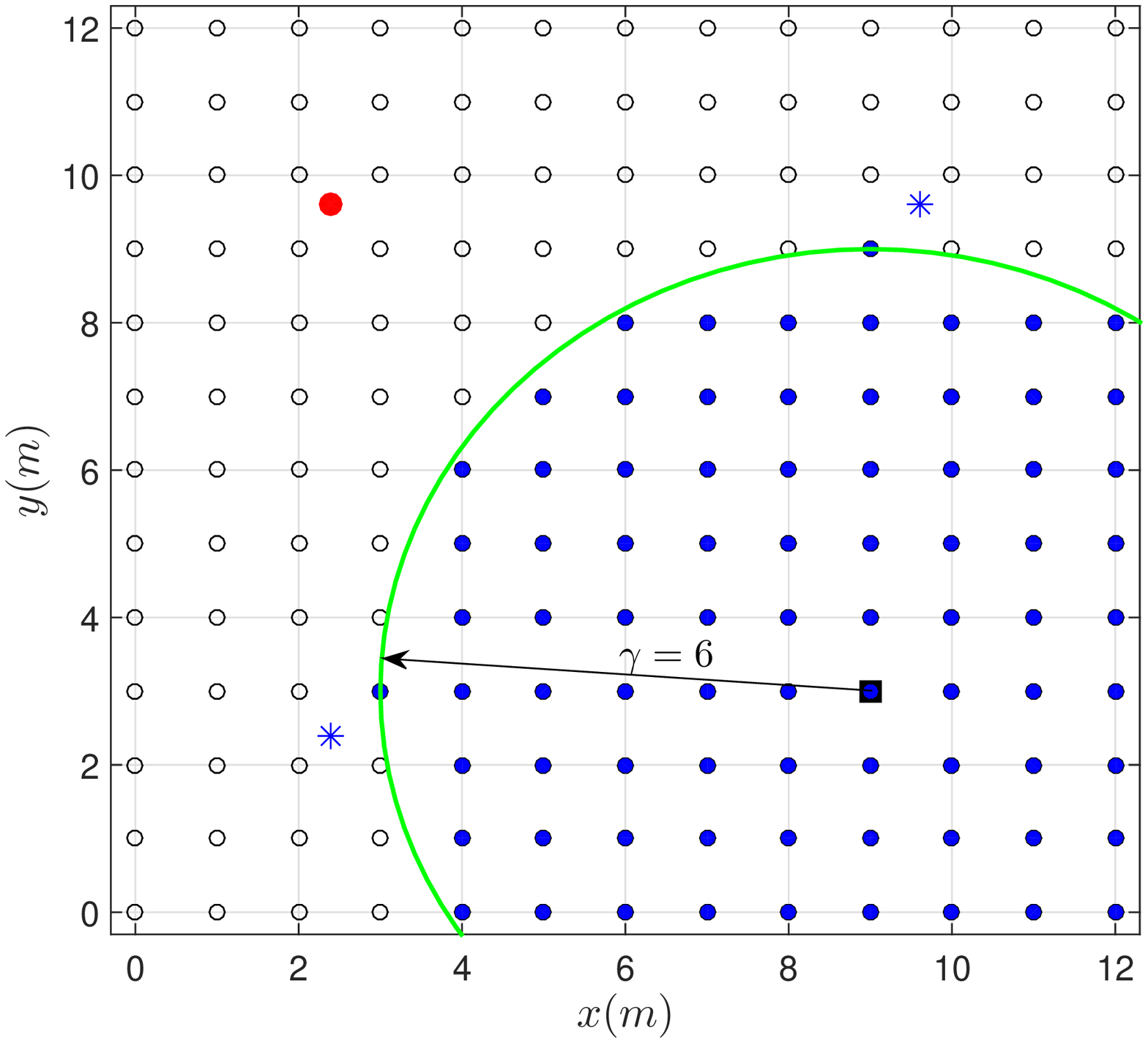}
      \hspace*{-2em}
      \vspace{-0.9cm}
     \label{fig:example_radius_selection}}
\end{subfigure}
  \begin{subfigure}[Sparse MVDR ($\mu=3.5\times 10^{-6}$)]
      {\includegraphics[width=0.35\textwidth, height=0.3\textwidth]{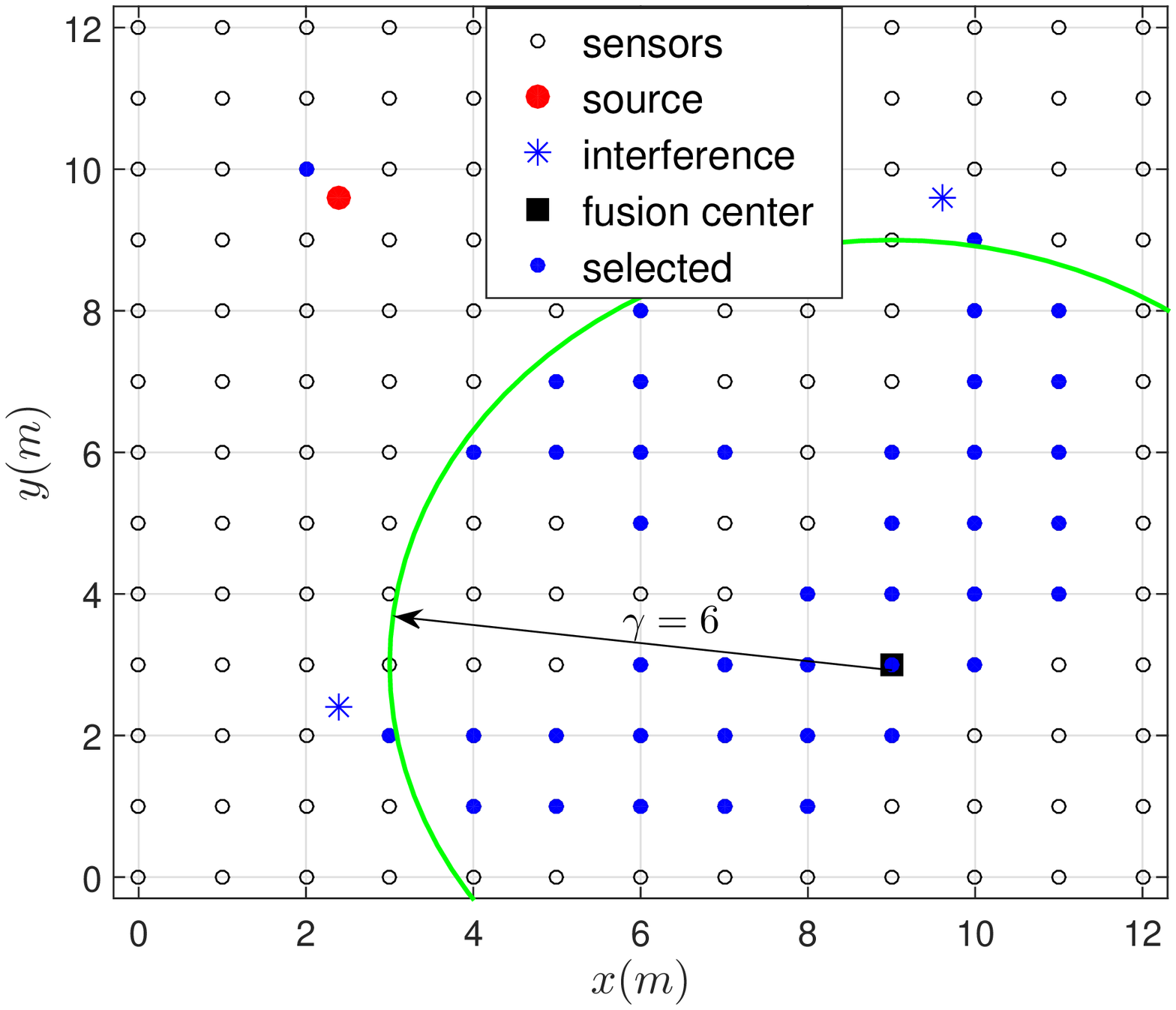}
      \hspace*{-2em}
      \vspace{-0.9cm}
      \label{fig:example_sparse_selection}}
  \end{subfigure}
    \begin{subfigure}[Backward selection ($c_T=0.09$)]
      {\includegraphics[width=0.35\textwidth, height=0.3\textwidth]{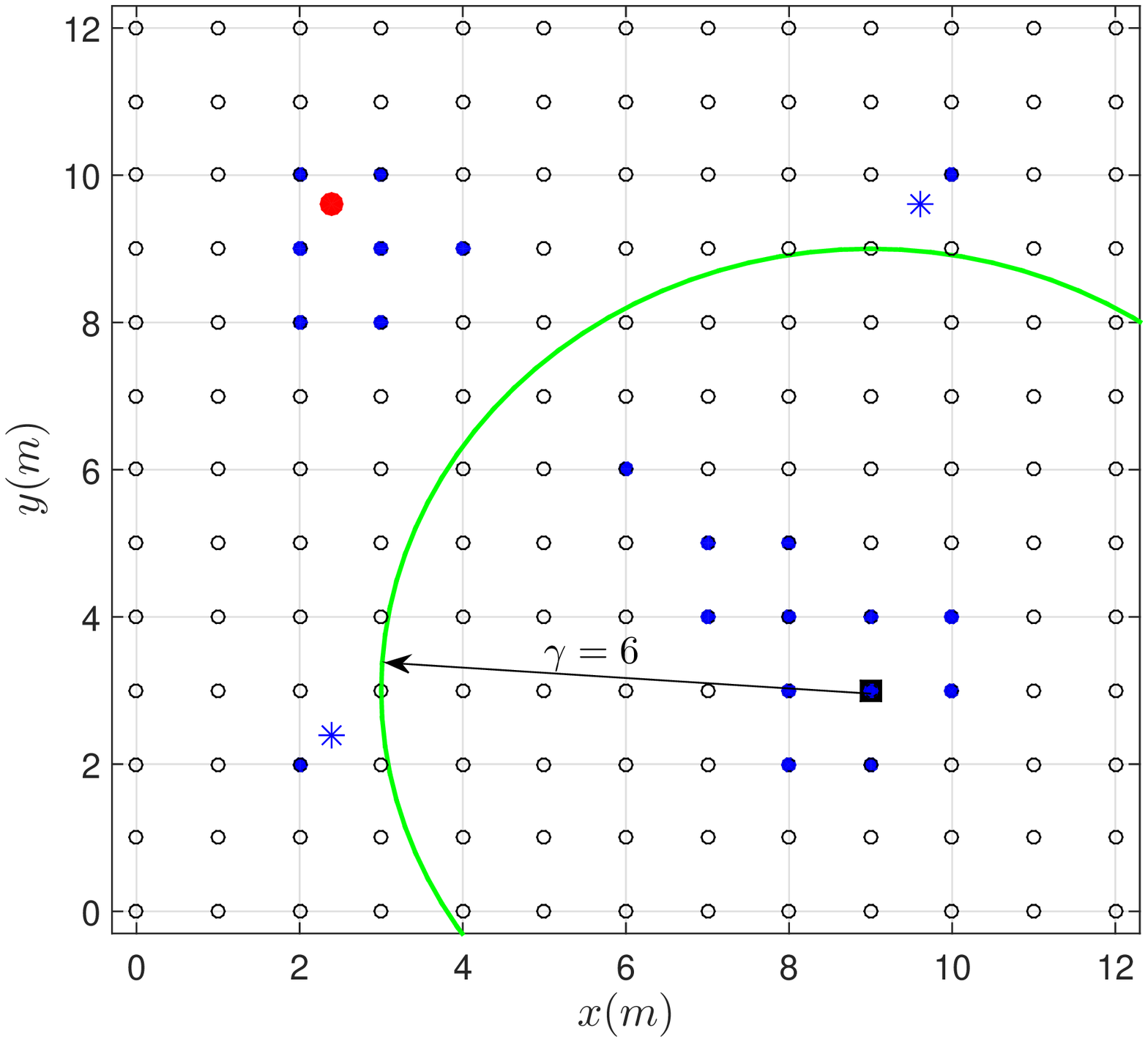}
      \hspace*{-2em}
      \vspace{-0.9cm}
      \label{fig:example_remove_selection}}
  \end{subfigure}\\
  \begin{subfigure}[Proposed approach ($\alpha=0.65$)]
      {\includegraphics[width=0.35\textwidth, height=0.3\textwidth]{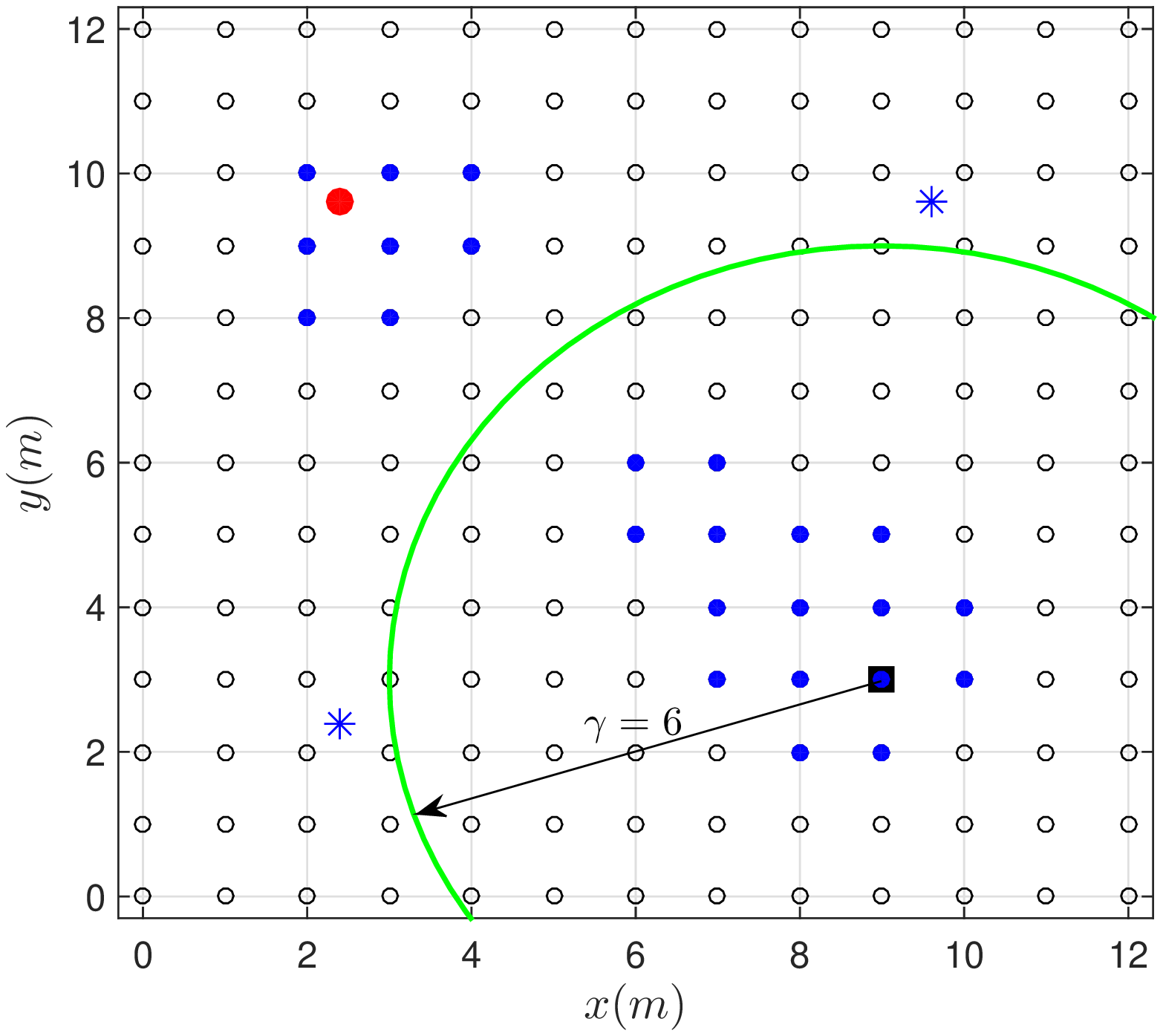}
      \hspace*{-2em}
      \vspace{-0.9cm}
      \label{fig:example_mintr_selection_alpha_0.65}}
  \end{subfigure}
    \begin{subfigure}[Proposed approach ($\alpha=0.9$)]
      {\includegraphics[width=0.35\textwidth, height=0.3\textwidth]{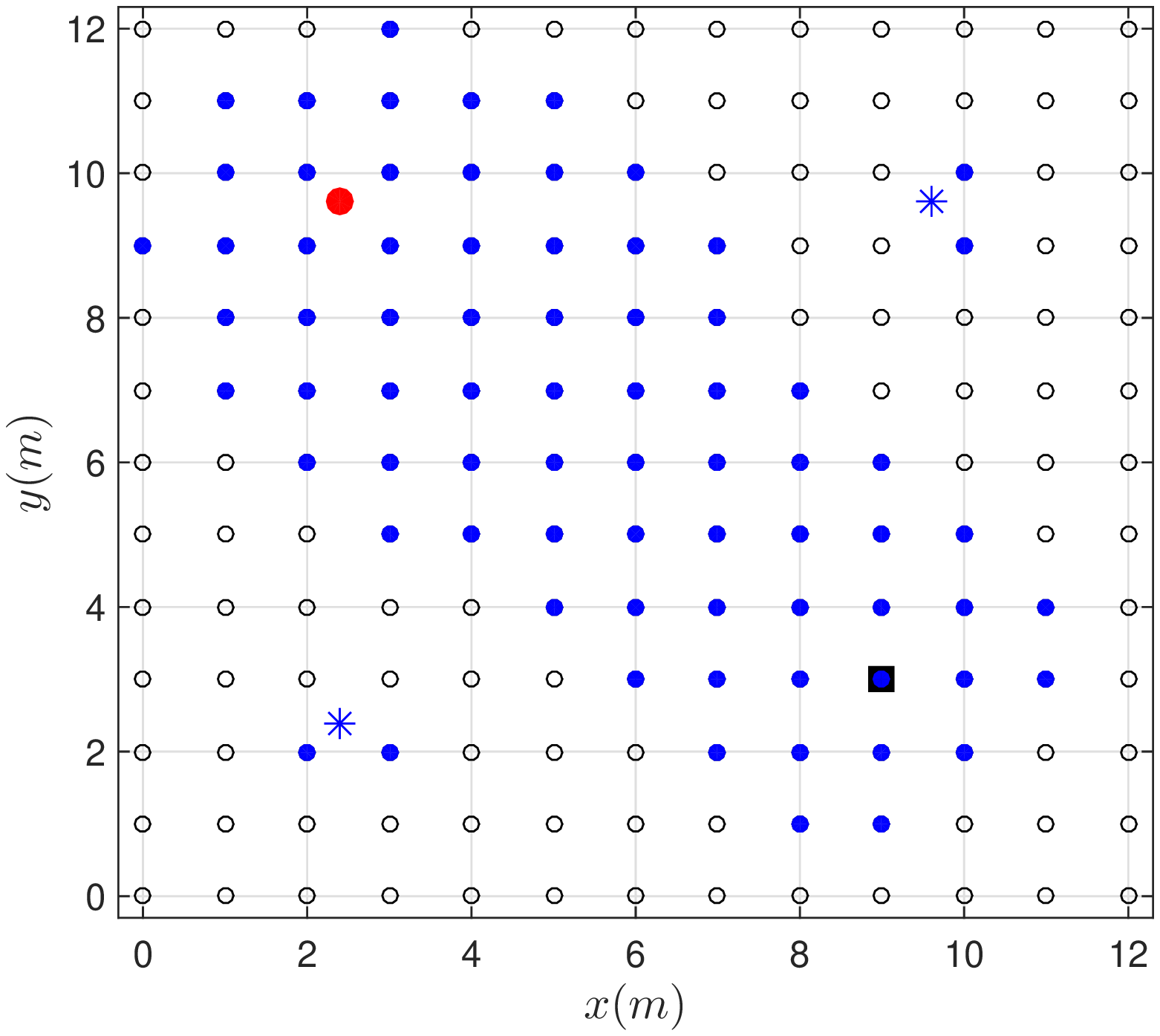}
      \hspace*{-2em}
      \label{fig:example_mintr_selection_alpha_0.9}}
  \end{subfigure}
  \begin{subfigure}[Uncorrelated case ($\alpha=0.9$)]
      {\includegraphics[width=0.35\textwidth, height=0.3\textwidth]{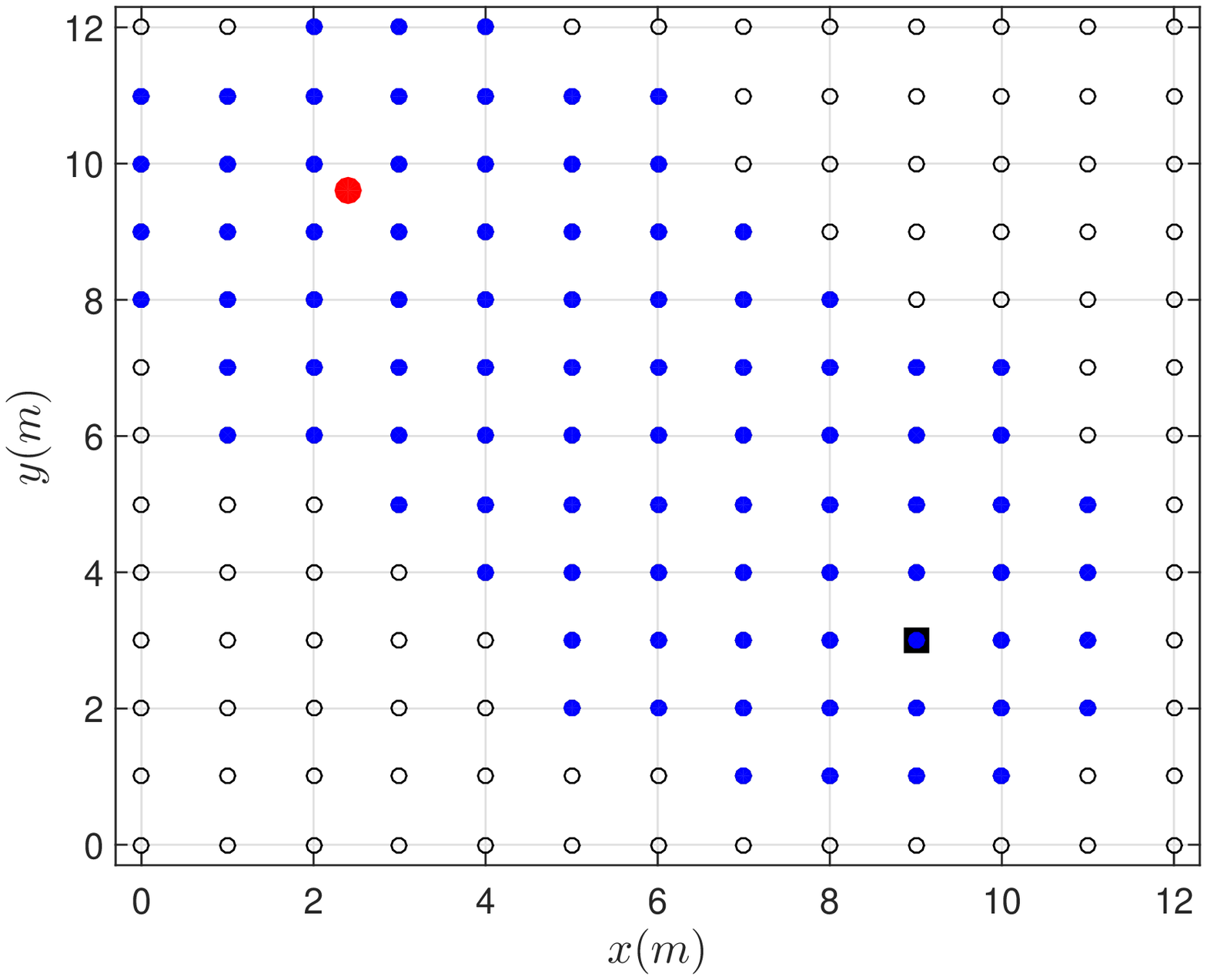}
      \hspace*{-2em}
      \label{fig:example_mintr_selection_uncorr}}
  \end{subfigure}
\caption{Microphone subset selection examples (The blue sensors  are activated for the MVDR beamformer): (a) radius-based MVDR beamforming, (b) sparse MVDR beamforming, (c) backward selection~[20] and (d) proposed method ($\alpha=0.65$)  for spatially correlated noises, respectively, (e) proposed method for correlated case with $\alpha=0.9$, and (f) proposed method ($\alpha=0.9$) for spatially uncorrelated noises only.}
\vspace{-0.3cm}
\label{fig:examples_model_based}
\end{figure*}
In order to compare the state-of-the-art approaches mentioned in Sec.~\ref{sec:refer_method}, we first investigate the influence of the required parameters $\alpha,\mu, \gamma, c_T$ on the performance, for the proposed and the different reference methods. Fig.~\ref{fig:alpha_mu_radius} shows the relationship between the output noise power (in dB) and the transmission power for $\text{SIR}=0$ dB with SIR representing signal-to-interference ratio. Fig.~\ref{fig:alpha_mu_radius} also shows the results without randomized rounding regarded as the lower bound of the proposed method, i.e., involving the selection variable $\mathbf{p}$ (thus, no selection) for computations. We can conclude that in order to reach the same noise reduction performance, the proposed approach always requires significantly less transmission costs compared to the weighted sparse beamformer or radius-based beamformer. If the transmission power budget $c_T$ (defined in Algorithm 2) is small, the proposed method performs better than the backward selection~[20], and if $c_T$ is large, they are comparable.   Furthermore, when $\alpha=0.65, \gamma=6,\mu=3.5\times 10^{-6}$, the four approaches approximately have the same transmission power as $c_T=0.09$. Hence, in the simulations that will follow we will compare the cases for $\alpha=0.65, \gamma=6,\mu=3.5\times 10^{-6},c_T=0.09$.

Fig.~\ref{fig:examples_model_based}(a)-(d) illustrate typical sensor selection examples for one angular frequency ($\omega=\pi/256$ rad/s) of the radius-based MVDR beamformer ($\gamma=6$), sparse MVDR beamformer $(\mu=3.5\times 10^{-6})$, backward selection $(c_T=0.09)$ and the proposed method $(\alpha=0.65)$, respectively. In addition, we show the radius for the radius-based MVDR, where all the sensors within this radius are selected, and thus not depicted explicitly in Fig.~\ref{fig:examples_model_based}(b)-(d). For fixed sensor and source locations, it is observed that the selected sensors are the same for most frequency bins. The sensors within the green circles ($\gamma=6$) are selected by the radius-based method, which chooses the $\gamma$-closest sensors relative to the FC for the MVDR beamformer. It can be seen that in order to save transmission power as well as to reduce noise, the proposed approach selects some microphones close to the source and some close to the FC for computation, while the sparse MVDR beamformer or radius-based method do not have this property. Although the backward selection has this property, it performs somewhat worse in noise reduction, which can be seen in Fig.~\ref{fig:alpha_mu_radius}. On one hand, the signals recorded by the microphones close to the source position are degraded less by the interfering source, and they preserve the target source better. Those microphones are helpful for enhancing the target source. On the other hand, the microphones close to the FC require less transmission power to transmit data to the FC. They are selected as they hardly add to the total transmission costs. When we increase the adaptive factor $\alpha$, more sensors that are close to the interference positions are selected as well, because they carry information on the interfering sources as shown in Fig.~\ref{fig:example_mintr_selection_alpha_0.9}.

In  Fig.~\ref{fig:example_mintr_selection_uncorr}, we also illustrate the case where interfering sources are absent, and the microphone recordings are only degraded by the microphone self noise, taking the noise level SNR=50 dB. Compared to Fig.~\ref{fig:example_mintr_selection_alpha_0.9}, most selected microphones are the same, and they are more aggregate to the source position as well as to the FC. The difference is whether to select sensors that are close to the interferences. From this comparison, we can also conclude that the sensors that are close to the interference are useful for cancelling the correlated noise.



\subsection{Evaluation of the data-driven approach}\label{sec:exp:greedy-method}
In this subsection, we will evaluate the proposed greedy approach compared to the model-driven algorithm and the utility-based method. The advantages of the greedy algorithm will be demonstrated from three perspectives, i.e., convergence behaviour, initialization, and for a scenario with a moving FC. Note that for the greedy approach, its convergence behaviour depends on the initial point $z_0$ and the transmission range.
\subsubsection{Convergence behaviour}\label{sec:exp:greedy:converge_behaviour}
In order to analyze the convergence behaviour of the proposed greedy approach, the sensor network topology in this work is viewed as a grid topology, such that its transmission range $R_0$ is fixed to the distance between two neighboring microphone nodes. In this part, we take the initial point $z_0$ at the position (9, 3) m as an example to show the convergence behaviour of the greedy algorithm. The effect of the choice of $z_0$ will be looked into later in this section.

\begin{figure*}[!t]
\centering
  \begin{subfigure}
      {\includegraphics[width=0.267\textwidth]{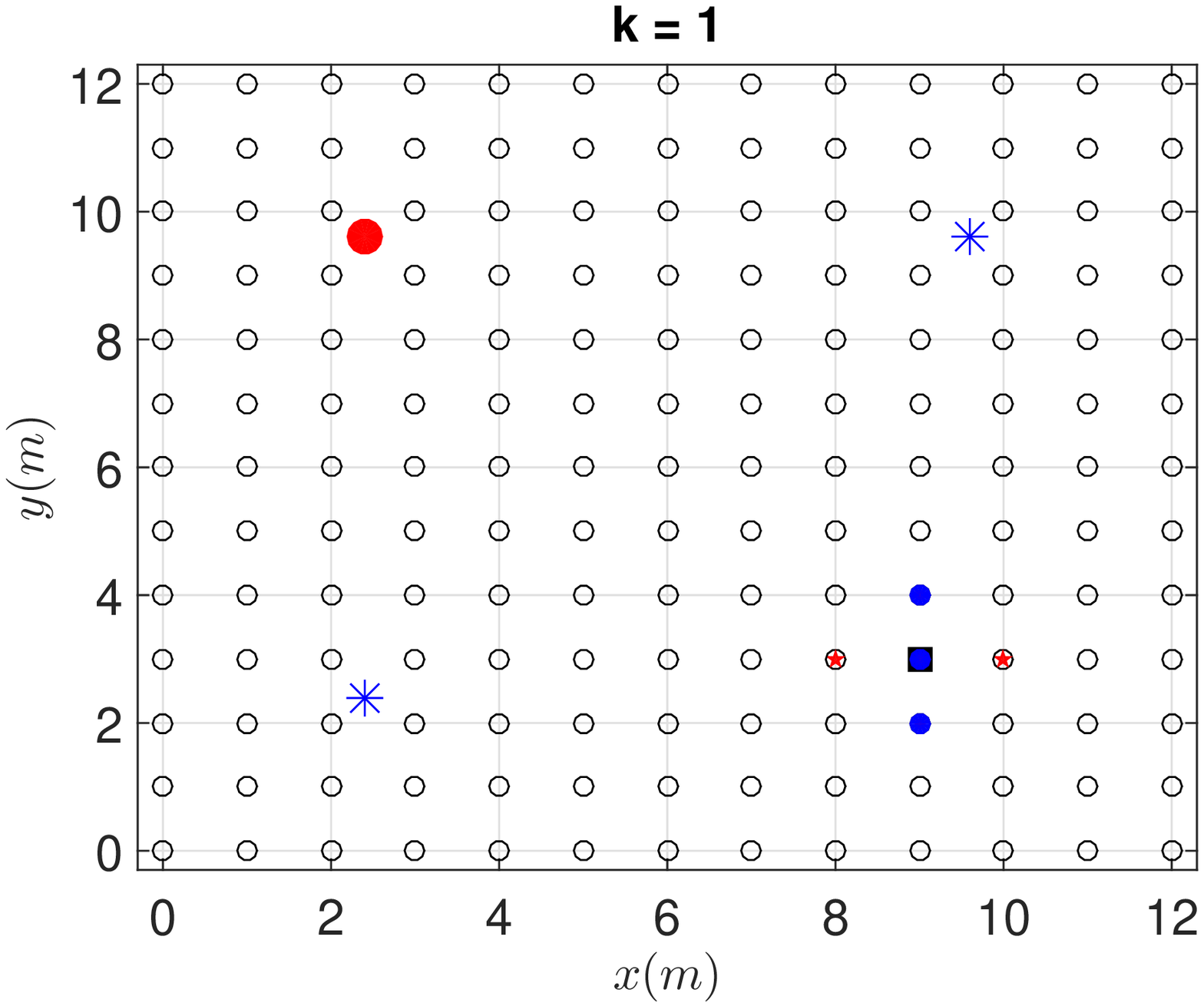}}
      \hspace*{-2em}
      \vspace{-0.1cm}
  \end{subfigure}
    \begin{subfigure}
      {\includegraphics[width=0.267\textwidth]{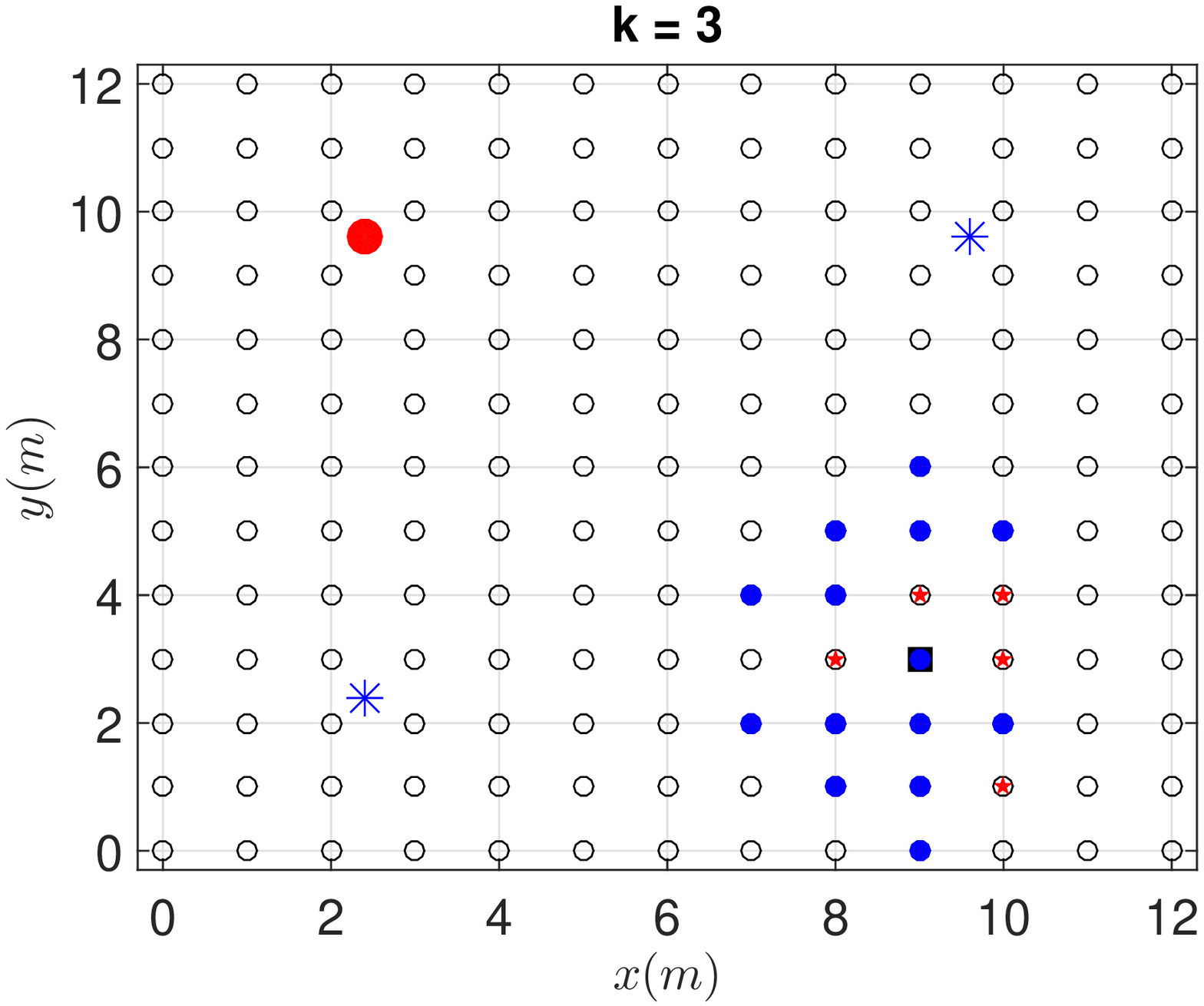}}
      \hspace*{-2em}
      \vspace{-0.1cm}
  \end{subfigure}
  \begin{subfigure}
      {\includegraphics[width=0.267\textwidth]{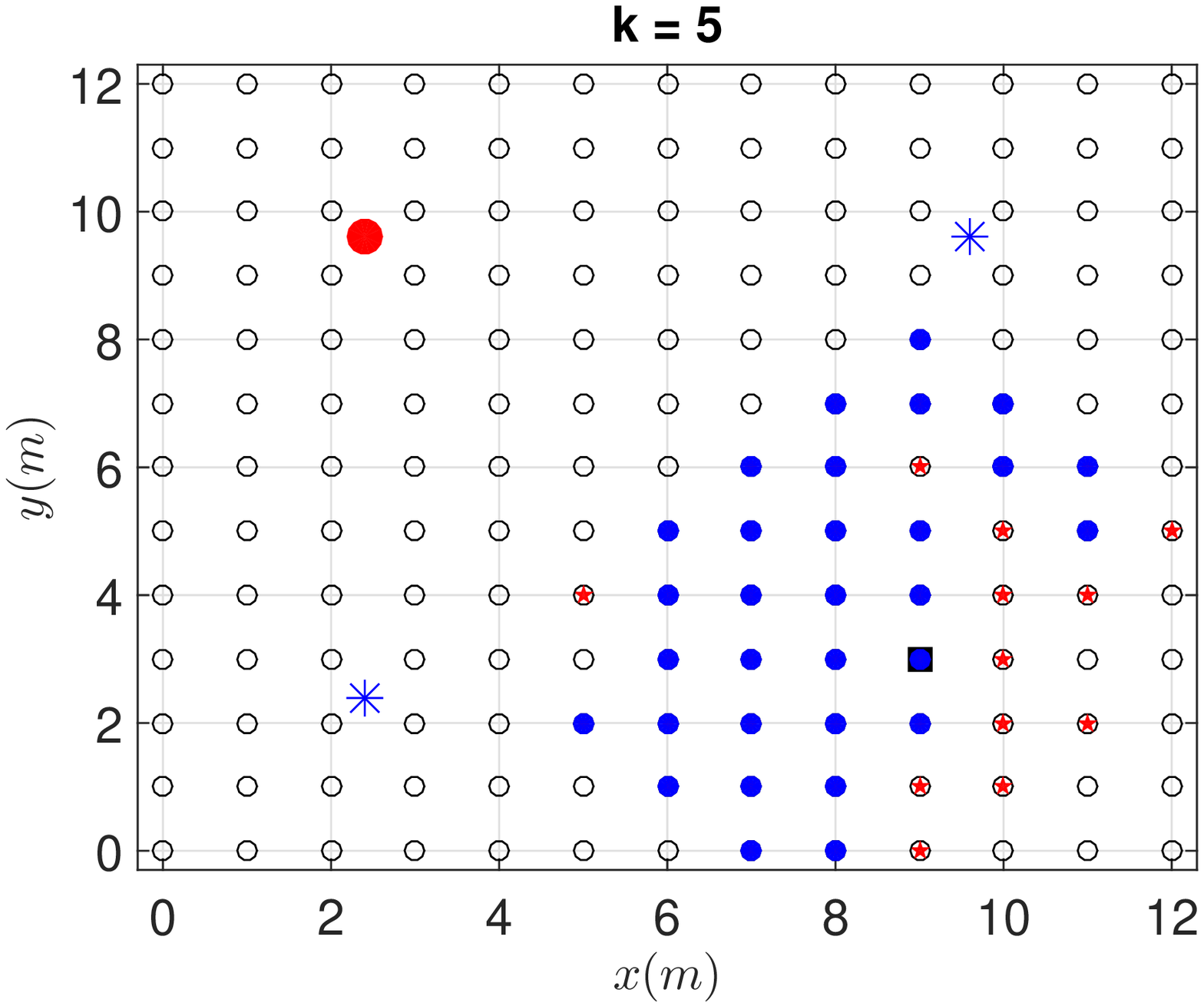}}
      \hspace*{-2em}
      \vspace{-0.1cm}
  \end{subfigure}
  \begin{subfigure}
      {\includegraphics[width=0.267\textwidth]{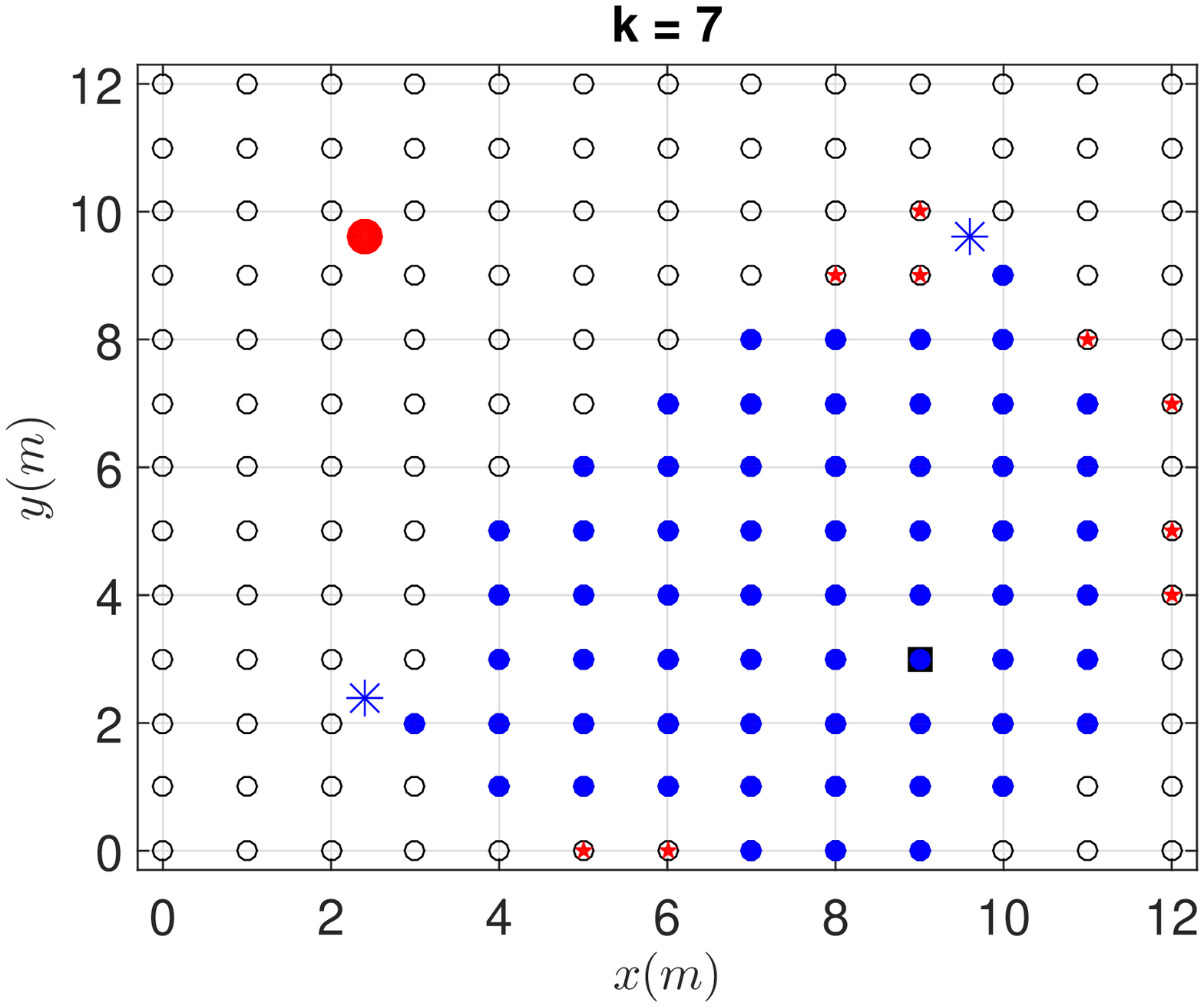}}
      \hspace*{-2em}
      \vspace{-0.1cm}
  \end{subfigure}\\
   \begin{subfigure}
      {\includegraphics[width=0.267\textwidth]{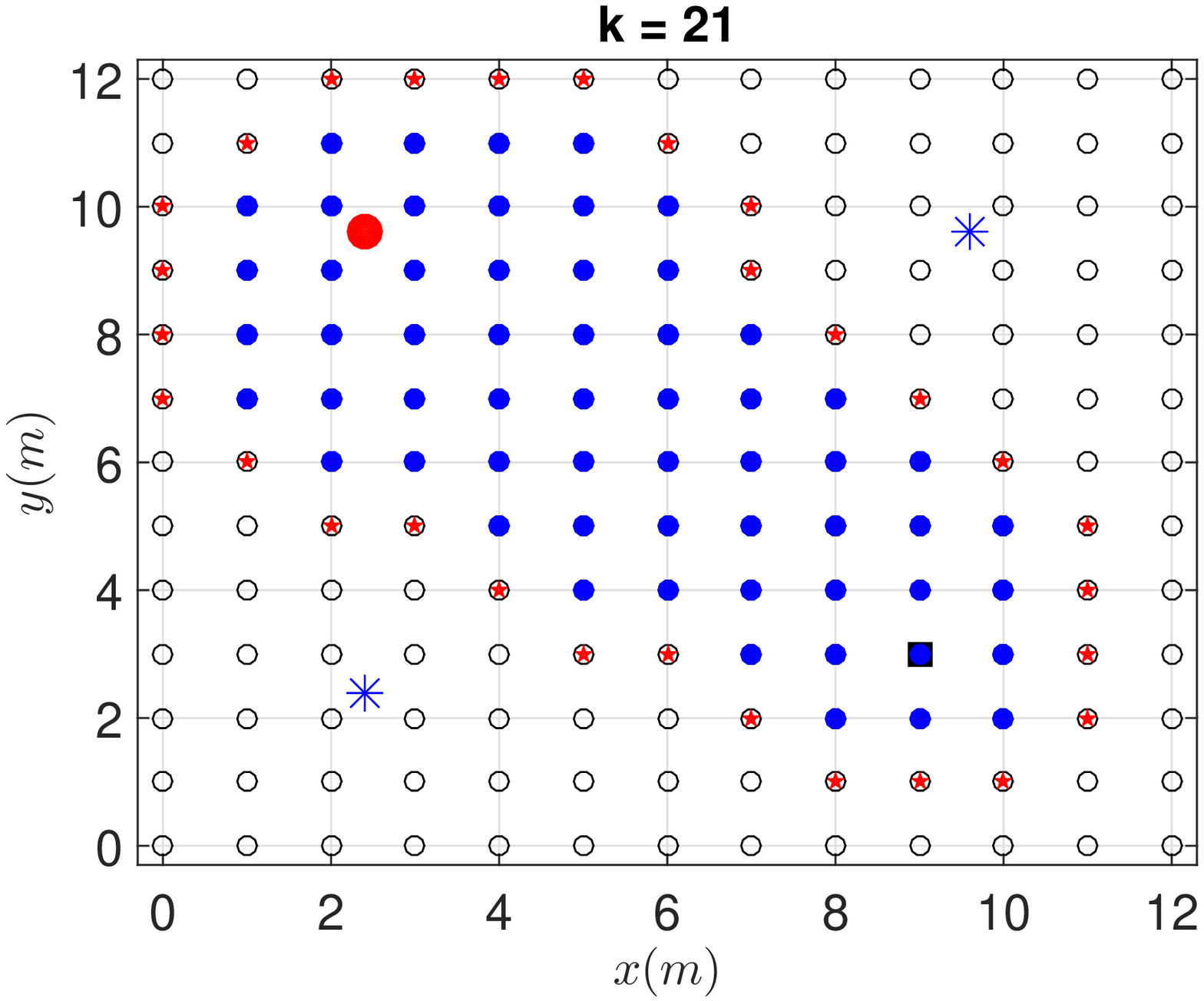}}
      \hspace*{-2em}
      \vspace{-0.1cm}
  \end{subfigure}
    \begin{subfigure}
      {\includegraphics[width=0.267\textwidth]{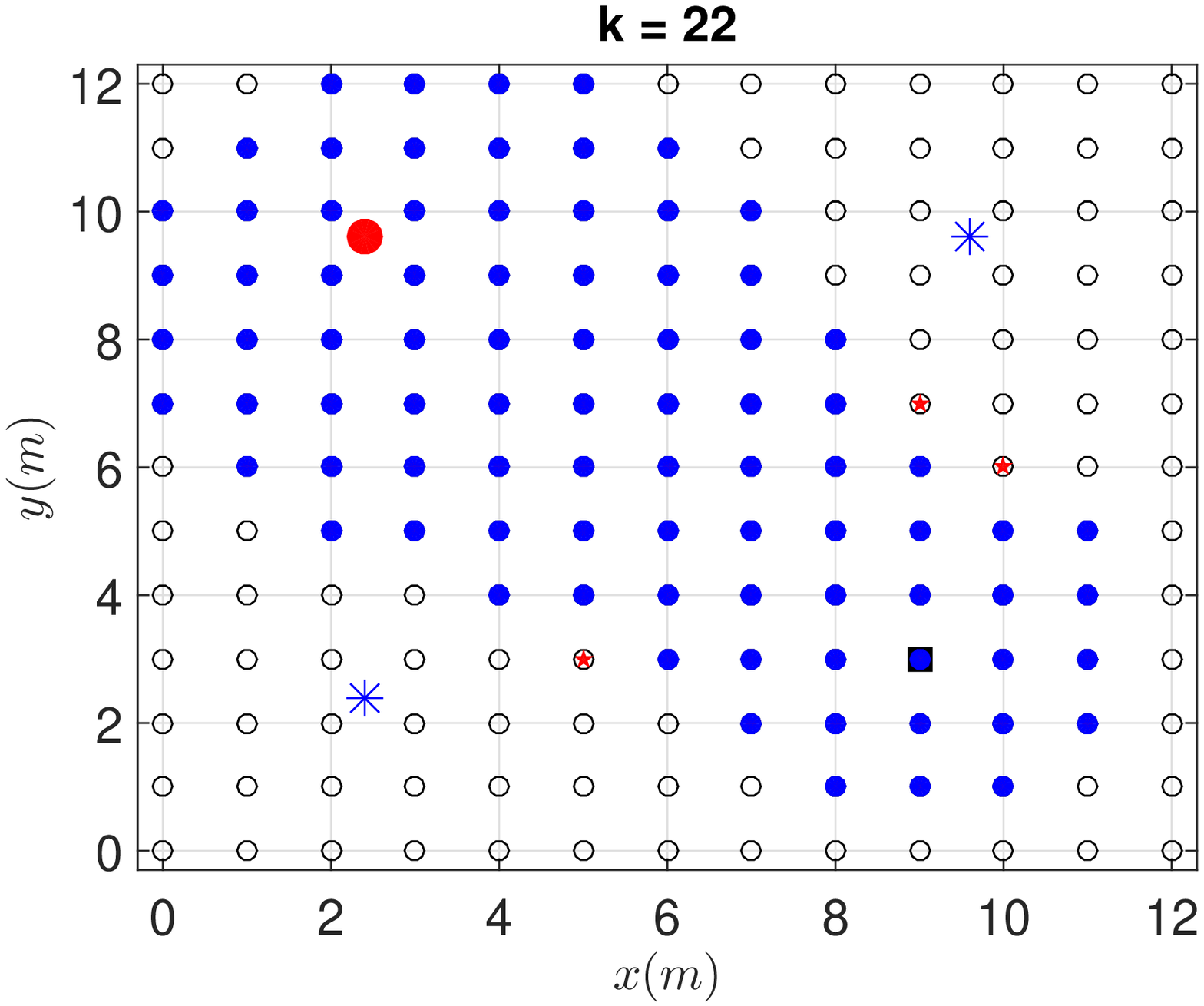}}
      \hspace*{-2em}
      \vspace{-0.1cm}
  \end{subfigure}
  \begin{subfigure}
      {\includegraphics[width=0.267\textwidth]{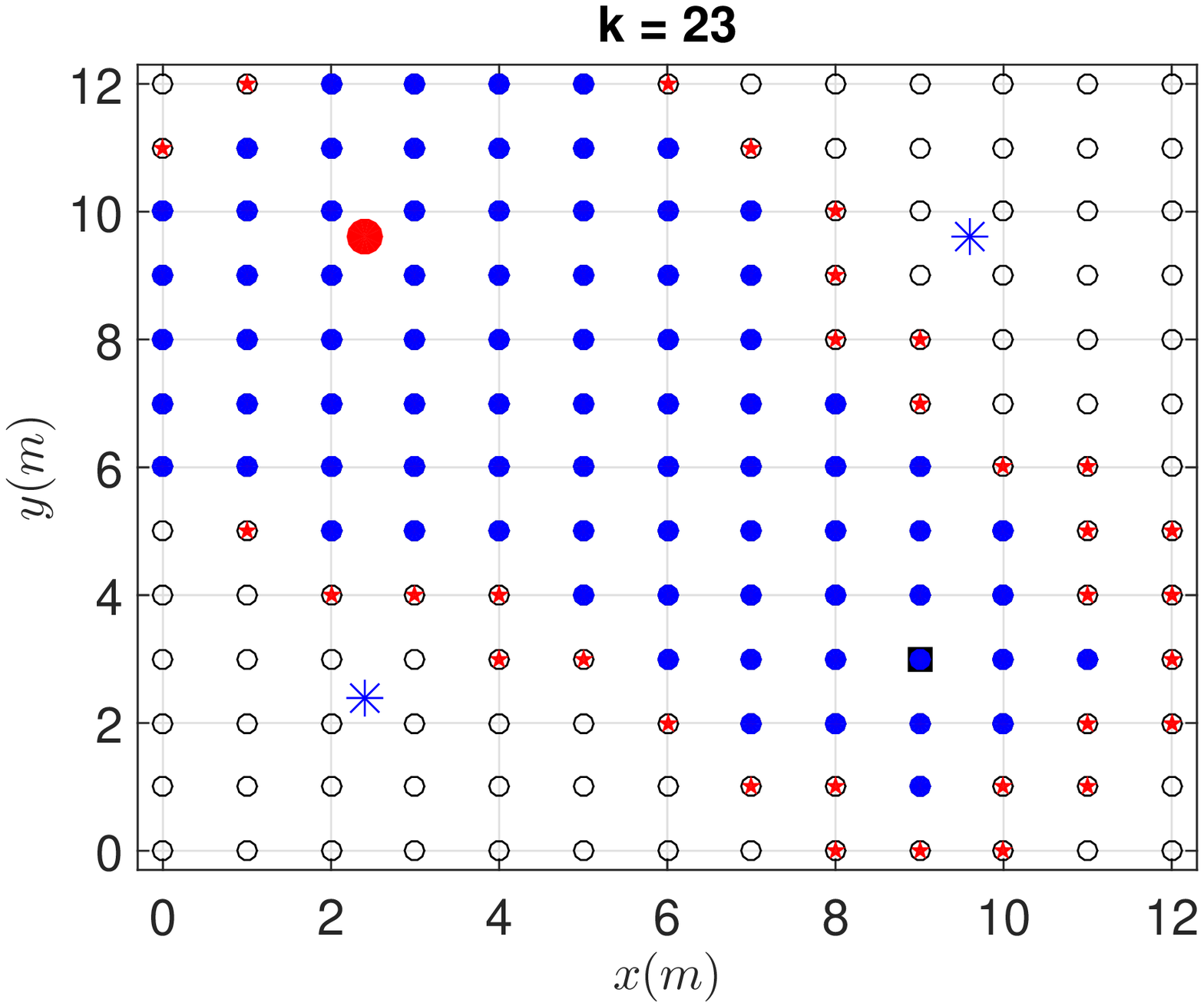}}
      \hspace*{-2em}
      \vspace{-0.1cm}
  \end{subfigure}
  \begin{subfigure}
      {\includegraphics[width=0.267\textwidth]{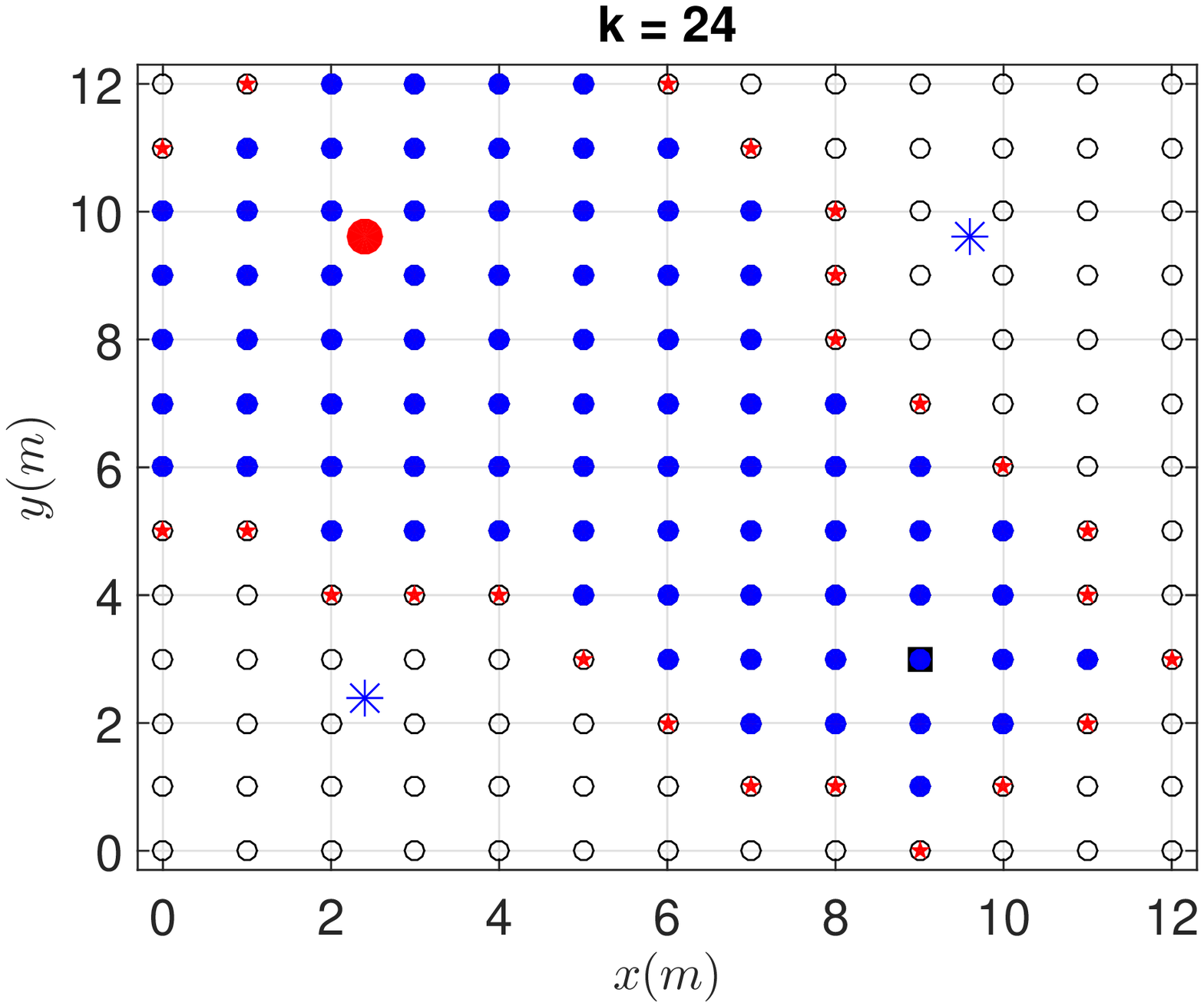}}
      \hspace*{-2em}
      \vspace{-0.1cm}
  \end{subfigure}\\
   \begin{subfigure}
      {\includegraphics[width=0.5\textwidth]{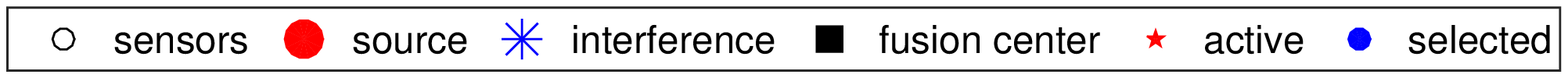}}
      \vspace{-0.5cm}
  \end{subfigure}
\caption{An illustration of the convergence behaviour for the proposed greedy algorithm (i.e., Algorithm 1). The initial point is located at (9, 3) m.}
\label{fig:greedy_local_global}
\vspace{-0.3cm}
\end{figure*}

Fig.~\ref{fig:greedy_local_global} illustrates the proposed greedy algorithm (i.e., Algorithm 1) for $\alpha=0.9$ using the same setup of Fig.~\ref{fig:example_mintr_selection_alpha_0.9}. In detail, at the 1st iteration (e.g., $k=1$) the $R_0$-closest candidate set $\mathcal{S}_1$ has five sensors. Based on the local constraint three sensors (in blue) are selected to form the set $\mathcal{S}_2$. The candidate set $\mathcal{S}_1$ is then increased by adding the $R_0$-closest sensors with respect to $\mathcal{S}_2$. This procedure continues for the first 21 iterations. When $k=21$, we can see that $\mathcal{S}_2$ is completely surrounded by $\mathcal{S}_1$, such that if we still use the local constraint, there would be no new sensors that can be added to $\mathcal{S}_1$, from which we conclude that the local constraint, i.e., $\beta_{\mathcal{S}_1}/\alpha$, has been satisfied. In order to satisfy the global constraint on the output noise power, the algorithm is then switched to the global constraint after the 21st iteration, i.e., $\beta/\alpha$. Finally, three more iterations are further required the reach the expected performance.

We can see from Fig.~\ref{fig:greedy_local_global}, that the proposed greedy method does not blindly increase the candidate set $\mathcal{S}_1$ towards all possible directions. Instead, $\mathcal{S}_1$ is increased only in the informative direction to the source location, such that the less informative microphones are not included. Furthermore, notice that the final selected set $\mathcal{S}_2$ differs slightly from the model-driven approach in Fig.~\ref{fig:example_mintr_selection_alpha_0.9}, as the greedy approach does not select the sensors that are close to the interfering sources, but it selects more sensors close to the target source. Hence, convergence towards the model-based approach is obtained in the sense of performance, but not in terms of selected sensors as the solution is not necessarily unique.

\begin{figure}
  \centering
  \includegraphics[width=0.4\textwidth]{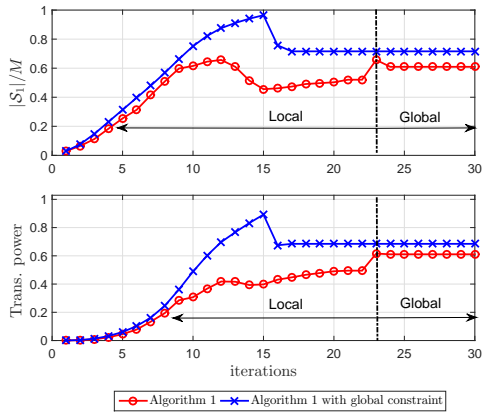}\\
  \vspace{-0.2cm}
  \caption{Cardinality of candidate set and transmission power vs iterations.}
  \label{fig:TP_numel_iterations}
  \vspace{-0.2cm}
\end{figure}

In Fig.~\ref{fig:TP_numel_iterations}, we show the ratio of cardinality of the candidate set $\mathcal{S}_1$ to the total number of sensors $M$  and transmission power per iteration. The combination of the global and local constraint is compared to a greedy algorithm that uses only the global constraint for Algorithm 1. Using only the global constraint,  $\mathcal{S}_1$ would blindly increase towards all directions. Clearly, we see that by using a combination between the local and the global constraint, much less sensors are included per iteration, such that the transmission power is kept low.

\subsubsection{Initializations}\label{sec:exp:initialization}
\begin{figure}[!t]
\centering
    \begin{subfigure}[centre (6, 6) m]
      {\includegraphics[width=0.25\textwidth]{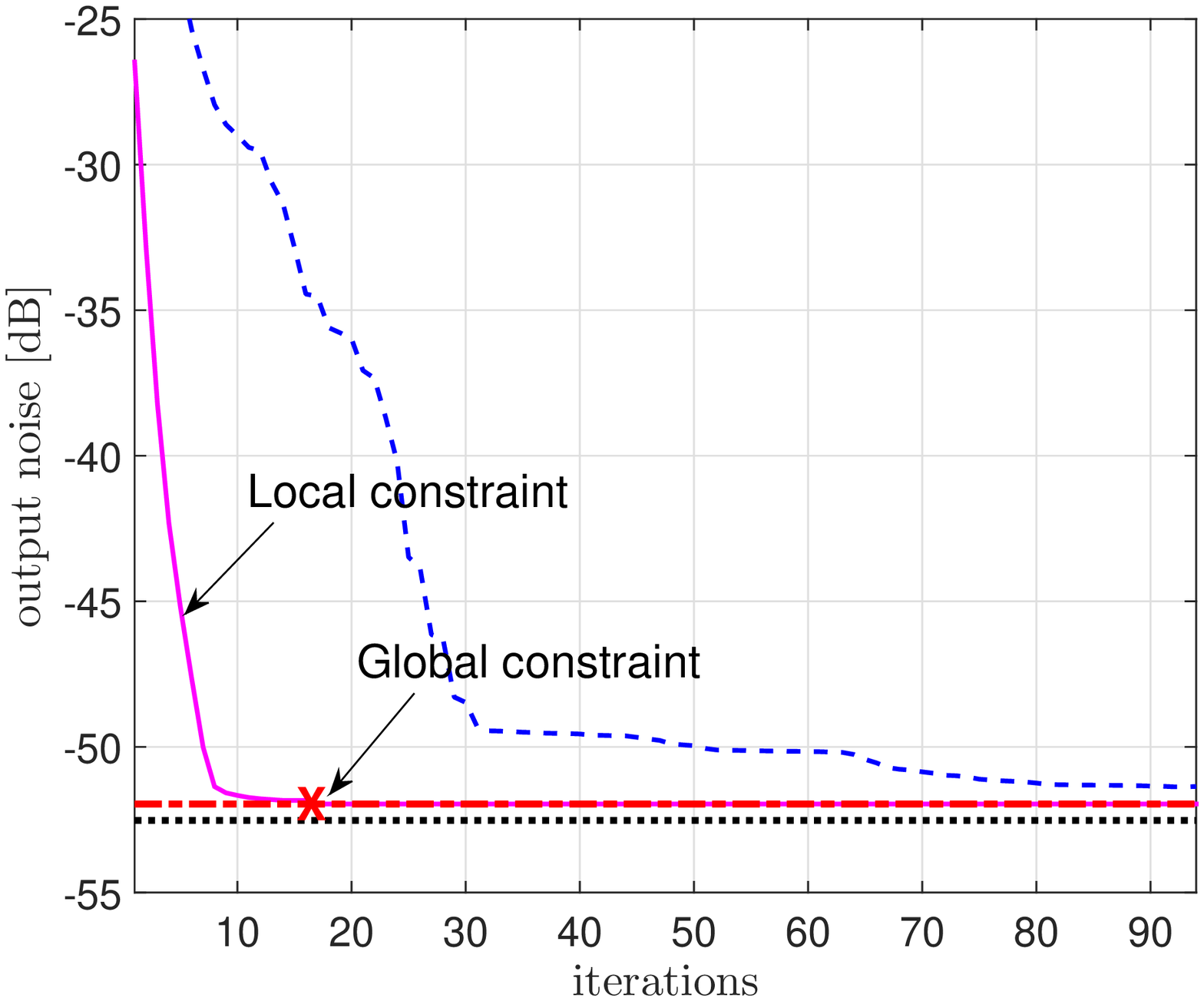}}
     \label{fig:init_greedy_npower_FC}
     \hspace*{-2em}
      \vspace{-0.1cm}
  \end{subfigure}
  \begin{subfigure}[source (2.4, 9.6) m]
      {\includegraphics[width=0.25\textwidth]{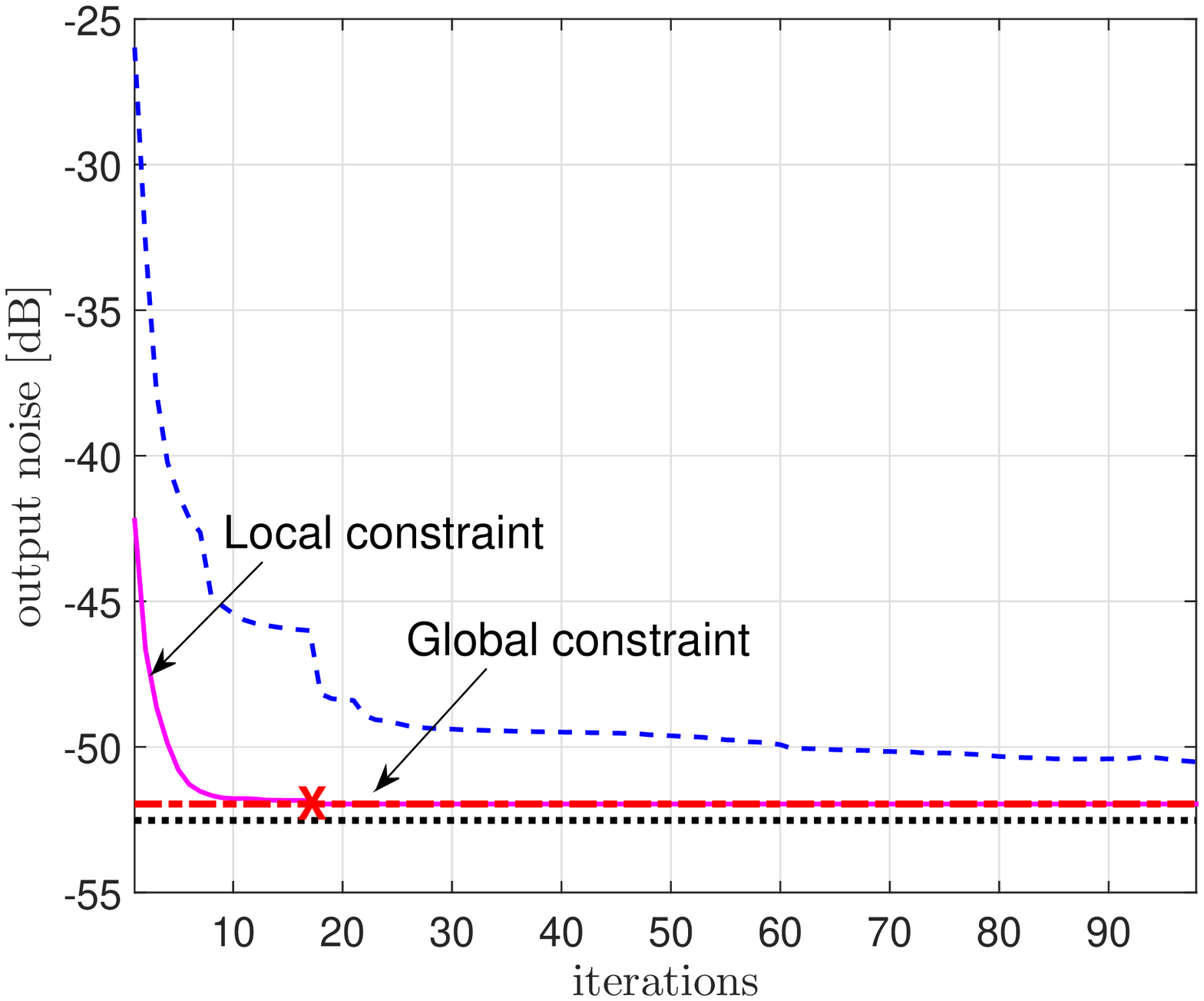}}
     \label{fig:init_greedy_npower_source}
     \hspace*{-2em}
      \vspace{-0.1cm}
  \end{subfigure}\\
  \begin{subfigure}[\#1 interference (2.4, 2.4) m]
      {\includegraphics[width=0.25\textwidth]{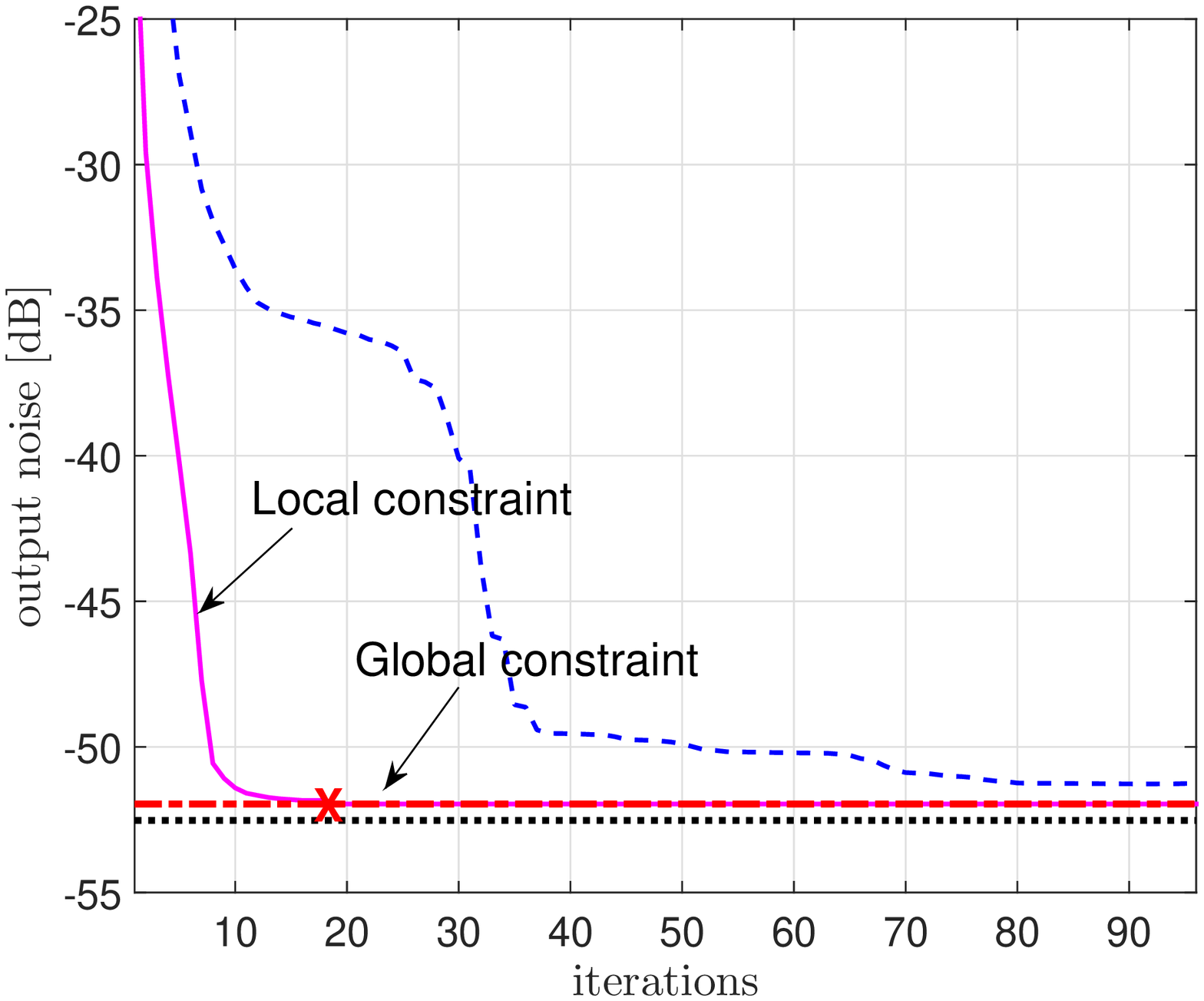}}
      \label{fig:init_greedy_npower_interference}
      \hspace*{-2em}
      \vspace{-0.1cm}
  \end{subfigure}
  \begin{subfigure}[FC (9, 3) m]
      {\includegraphics[width=0.25\textwidth]{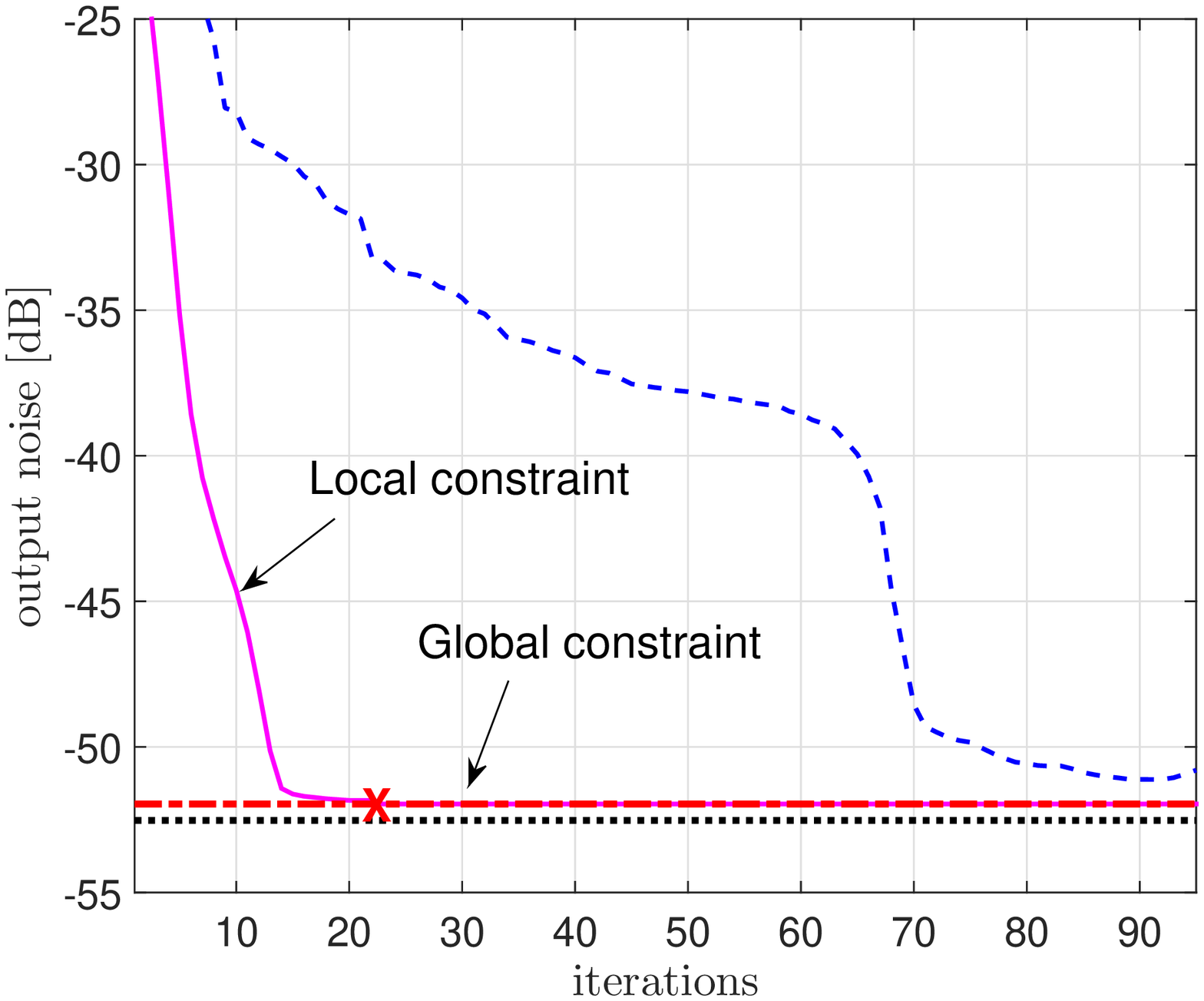}}
     \label{fig:init_greedy_npower_random}
     \hspace*{-2em}
      \vspace{-0.1cm}
  \end{subfigure}\\
     \begin{subfigure}
      {\includegraphics[width=0.5\textwidth]{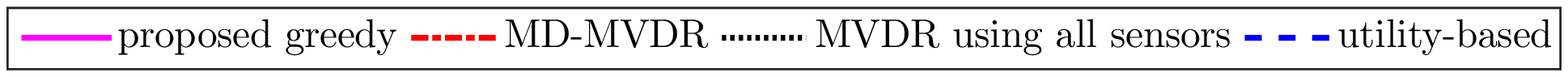}}
      \vspace{-1cm}
  \end{subfigure}
\caption{Output noise power in terms of iterations for different initial point $z_0$: (a) centre, (b) source, (c) interference, (d) FC.}
\label{fig:init_greedy_npower}
\vspace{-0.5cm}
\end{figure}

In this part, we will show the effect of the initial point $z_0$ on the convergence rate. Fig.~\ref{fig:init_greedy_npower} illustrates the output noise power (in dB) in terms of iterations for four different initializations, i.e., centre (6, 6) m, source position (2.4, 9.6) m, interference position (2.4, 2.4) m and FC (9, 3) m.
The red dashed line represents the performance of the model-driven algorithm proposed in Sec.~\ref{sec:model-selection}, which selects the most informative sensors from all the possible candidates. The black dashed line denotes the performance of the classical MVDR beamformer using all microphones. The magenta curve  shows the proposed greedy algorithm for the MVDR beamformer. The blue dashed curve denotes the performance of the utility-based algorithm~[19][20]. The output noise power of the greedy algorithm includes two steps: local constraint ($\beta_{\mathcal{S}_1}/\alpha$) and global constraint ($\beta/\alpha$). The moment that the constraint is switched from the local to the global constraint is indicated by the red marker ``$\times$". When executing the local constraint, the output noise power decreases fastest for the initialization at the source position and slowest for the FC initialization. This is due to the fact that the sensors that are close to the source are more informative for speech enhancement. After the algorithm converges based on the local constraint, by switching to the global constraint, the output noise decreases further until it reaches the performance of the model-driven approach.
Hence, from a perspective of performance, the proposed greedy algorithm converges to the model-driven method. In addition, if the point of initialization is closer to the source position, the convergence is faster. To conclude, the initialization only influences the convergence rate, and it does not affect the final performance.

Furthermore, from Fig.~\ref{fig:init_greedy_npower} we observe that the proposed greedy algorithm converges much faster than the utility-based method, because the latter only selects one sensor in each iteration. Note that in the comparisons the total transmission cost budgets for the two approaches are kept the same. Also, there is no guarantee for the utility-based method to fulfil the expected performance on the output noise power.
\subsubsection{Moving FC}
\begin{figure*}[!t]
\centering
  \begin{subfigure}
      {\includegraphics[width=0.267\textwidth]{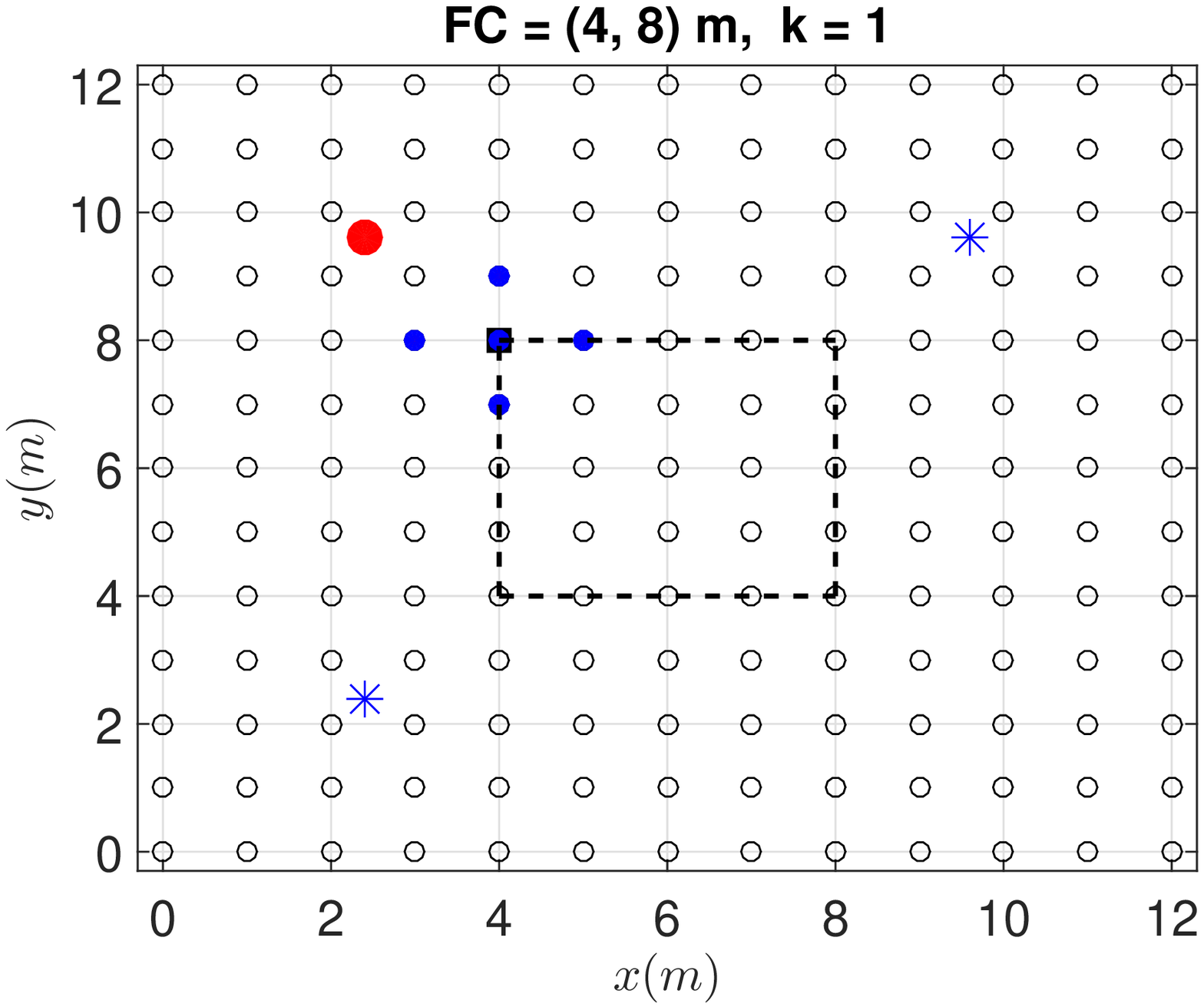}}
      \hspace*{-2em}
      \vspace{-0.05cm}
  \end{subfigure}
    \begin{subfigure}
      {\includegraphics[width=0.267\textwidth]{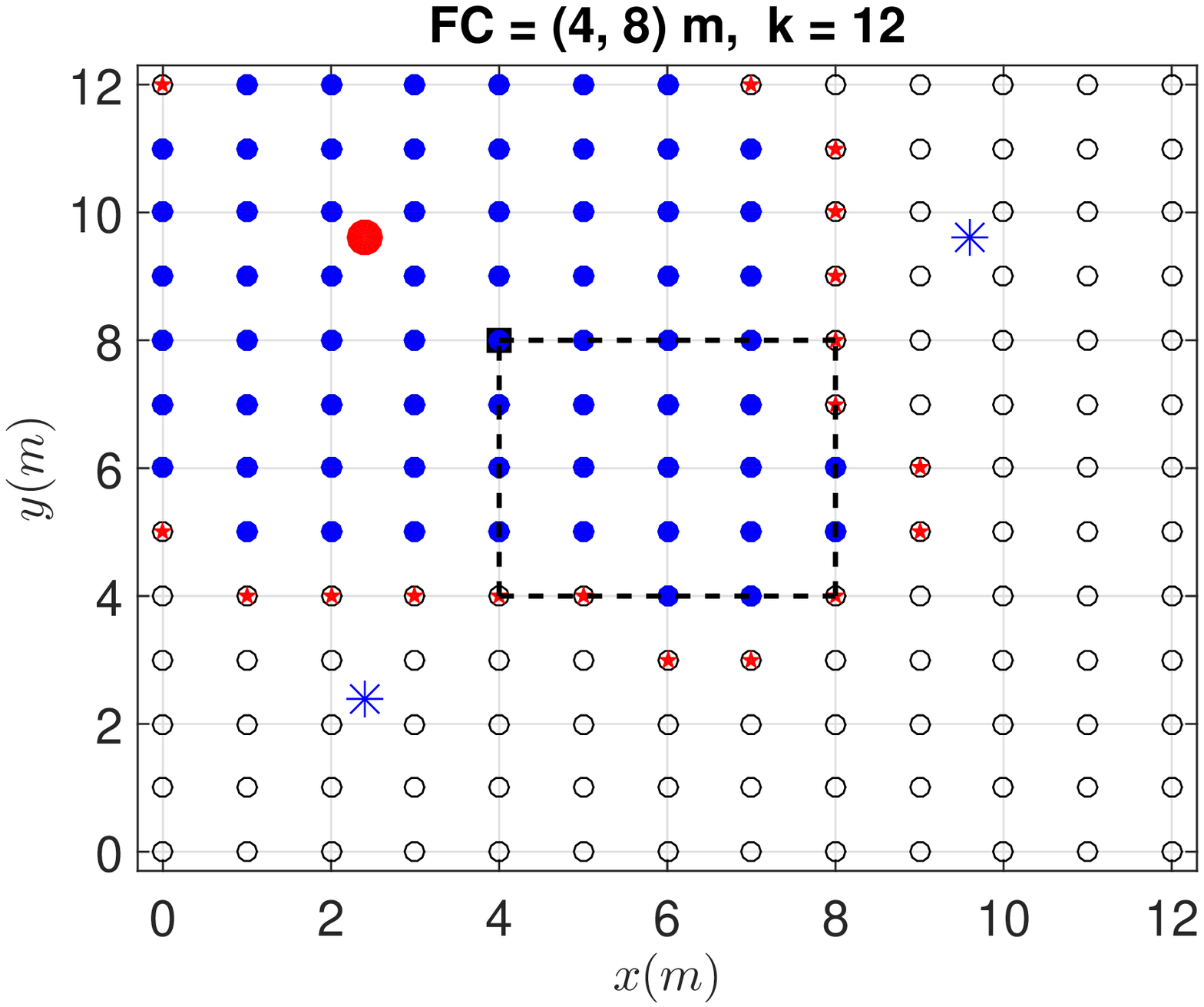}}
      \hspace*{-2em}
      \vspace{-0.05cm}
  \end{subfigure}
  \begin{subfigure}
      {\includegraphics[width=0.267\textwidth]{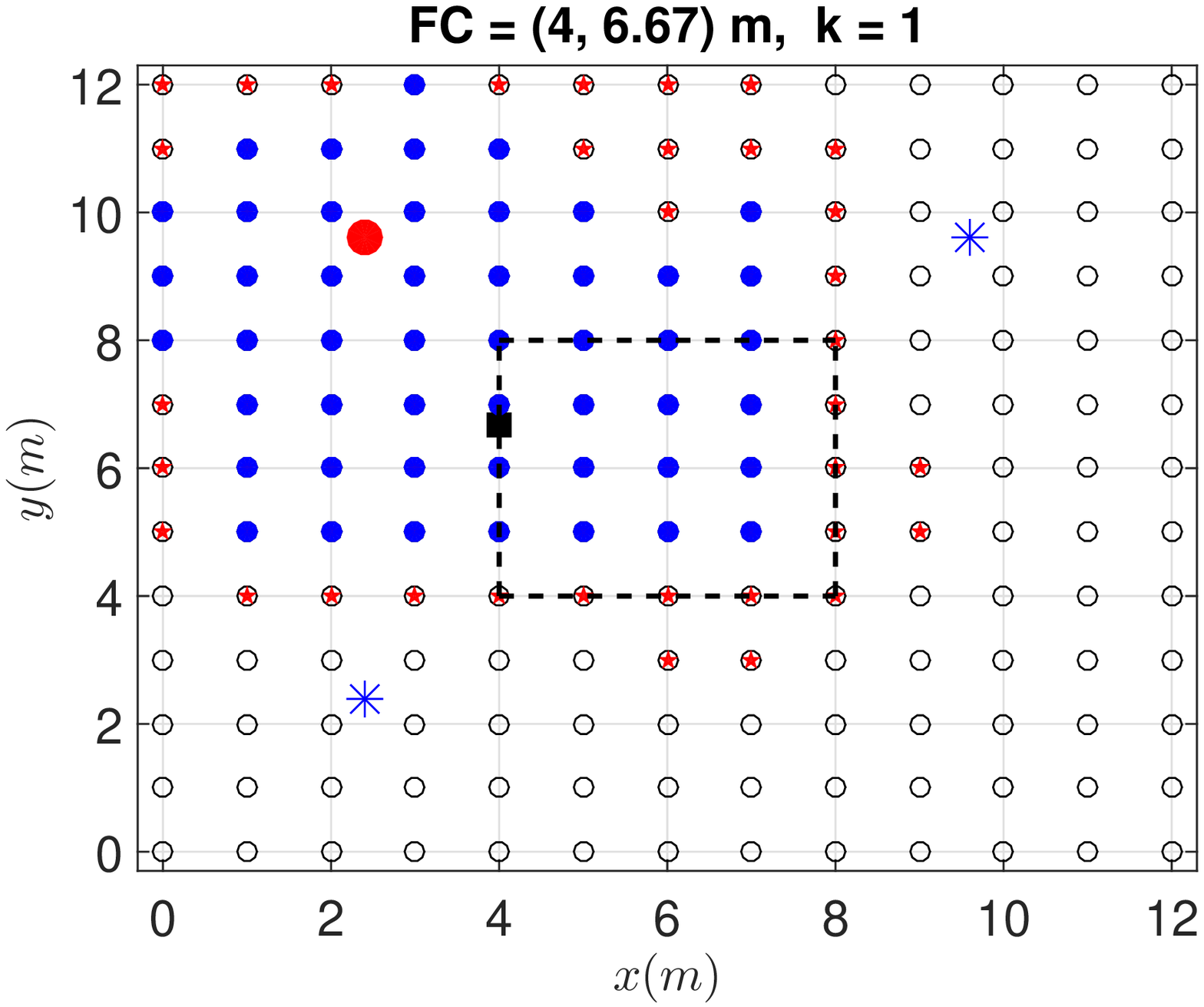}}
      \hspace*{-2em}
      \vspace{-0.05cm}
  \end{subfigure}
  \begin{subfigure}
      {\includegraphics[width=0.267\textwidth]{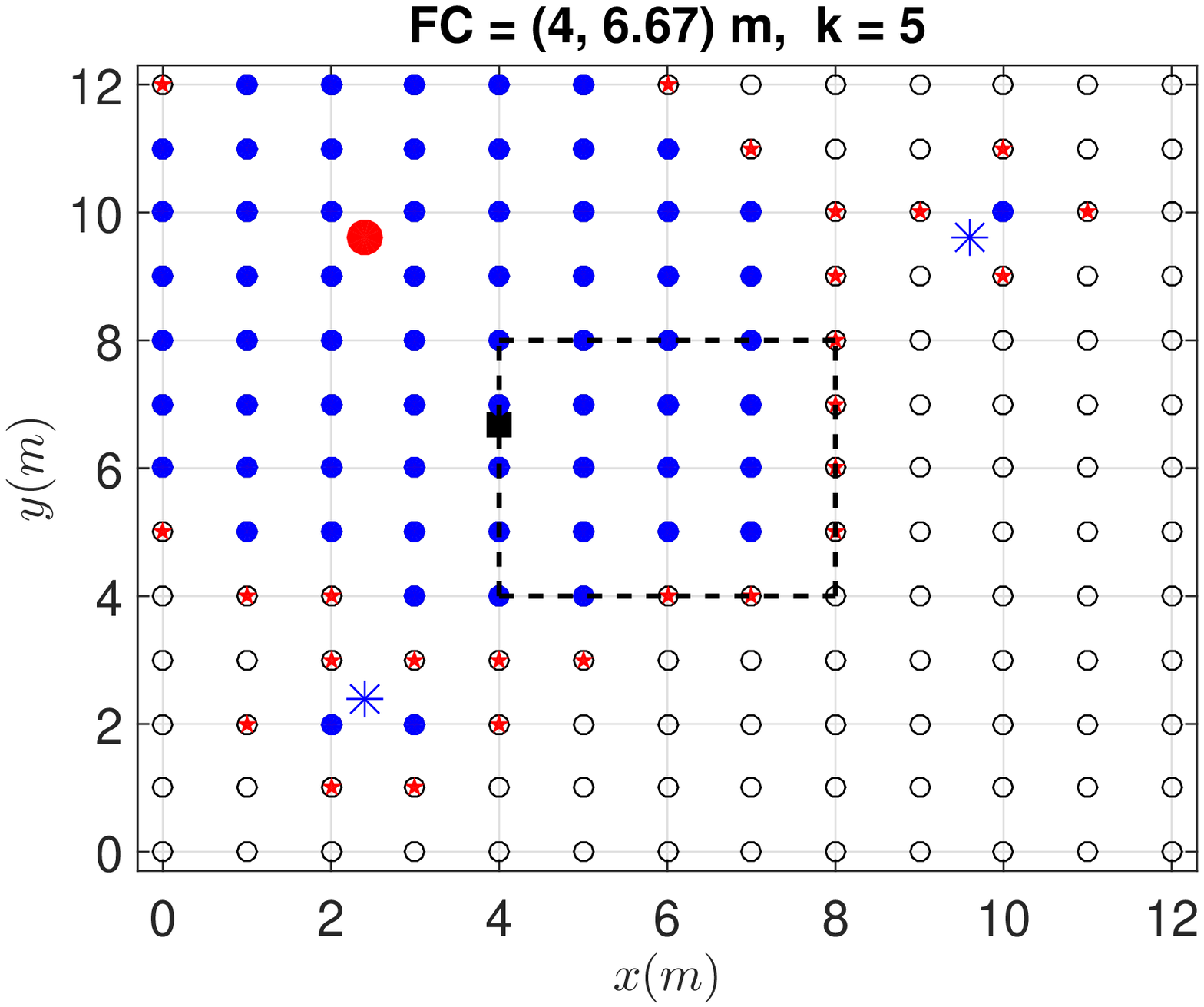}}
      \hspace*{-2em}
      \vspace{-0.05cm}
  \end{subfigure}\\
  \begin{subfigure}
      {\includegraphics[width=0.267\textwidth]{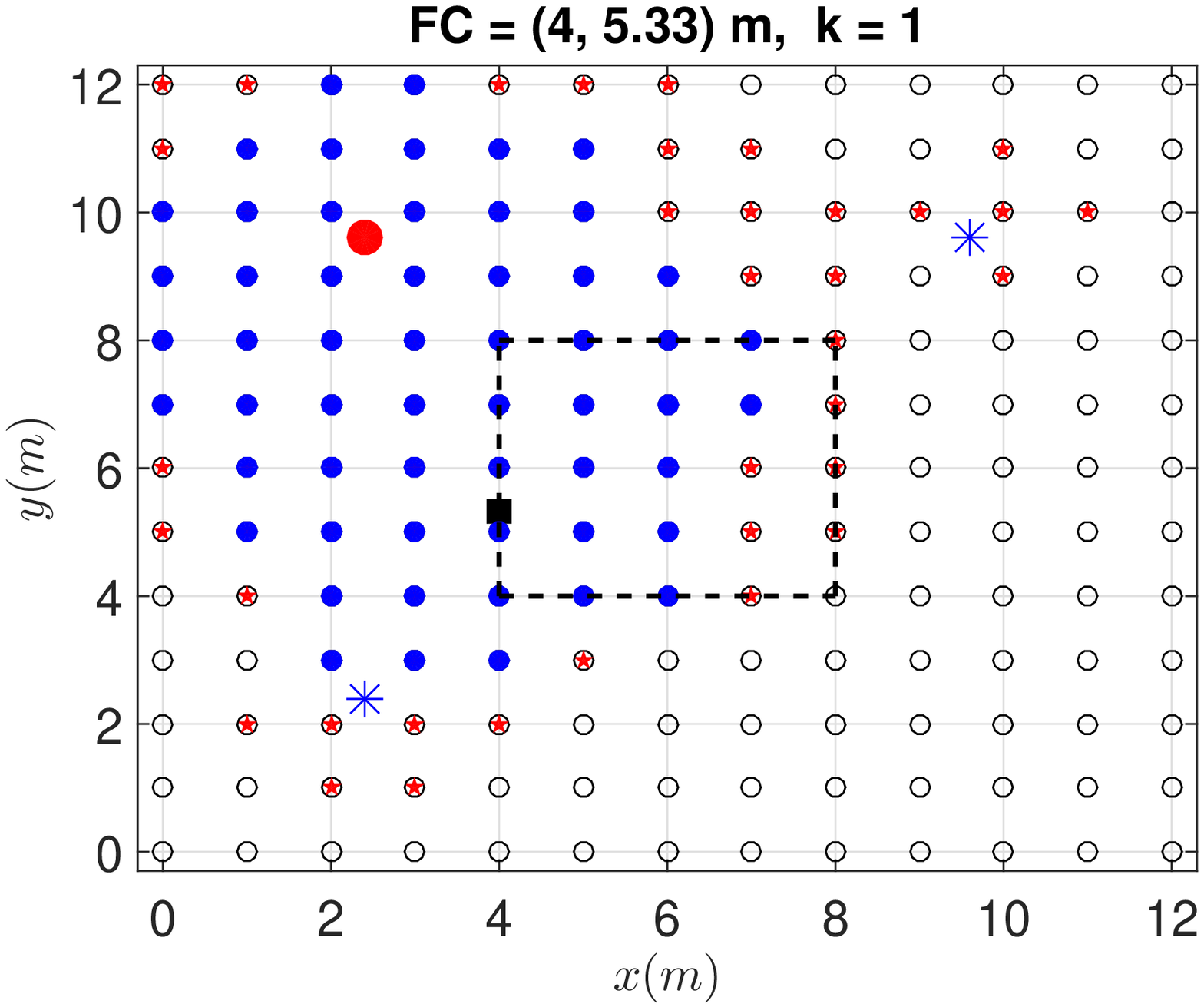}}
      \hspace*{-2em}
      \vspace{-0.1cm}
  \end{subfigure}
    \begin{subfigure}
      {\includegraphics[width=0.267\textwidth]{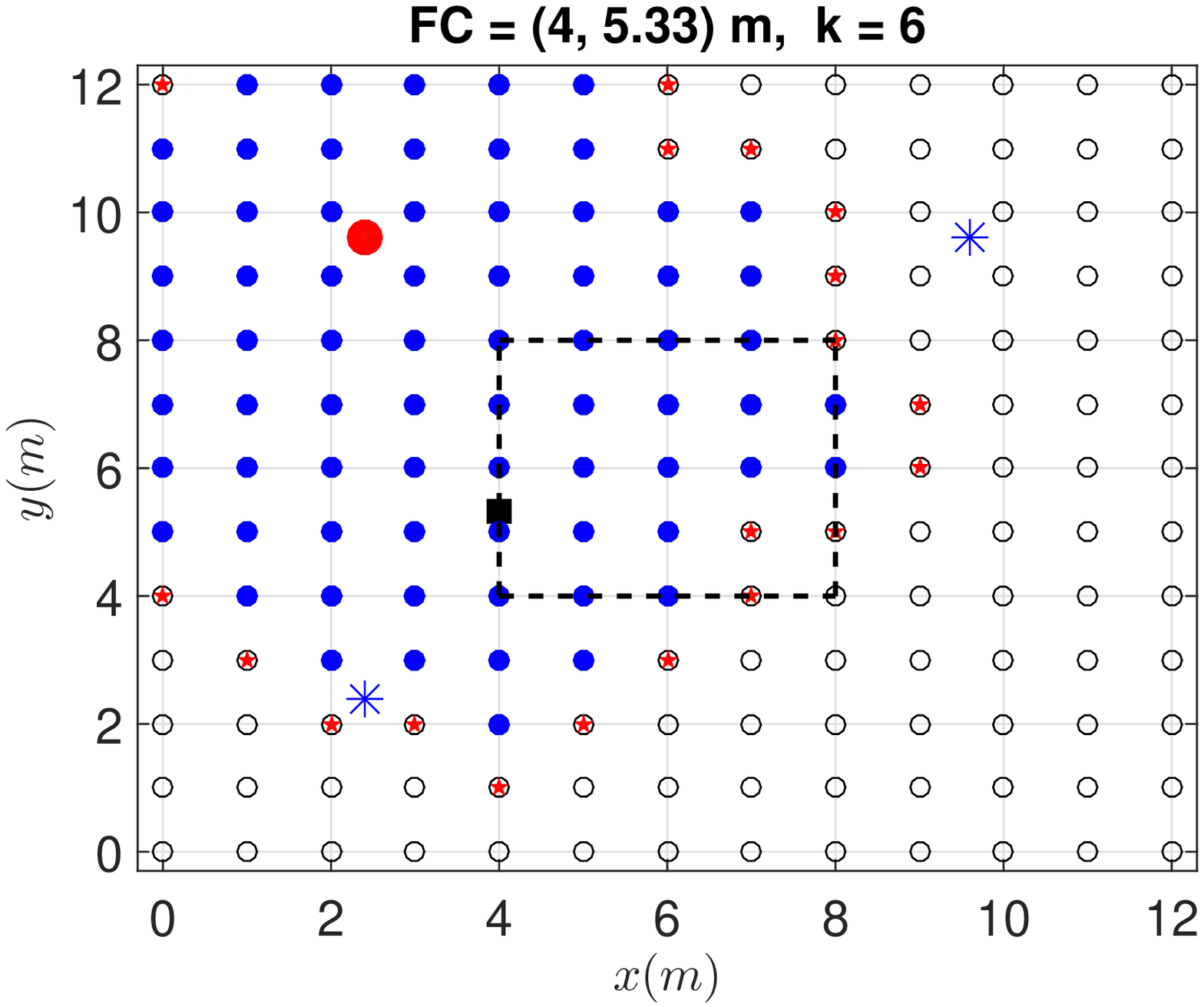}}
      \hspace*{-2em}
      \vspace{-0.1cm}
  \end{subfigure}
  \begin{subfigure}
      {\includegraphics[width=0.267\textwidth]{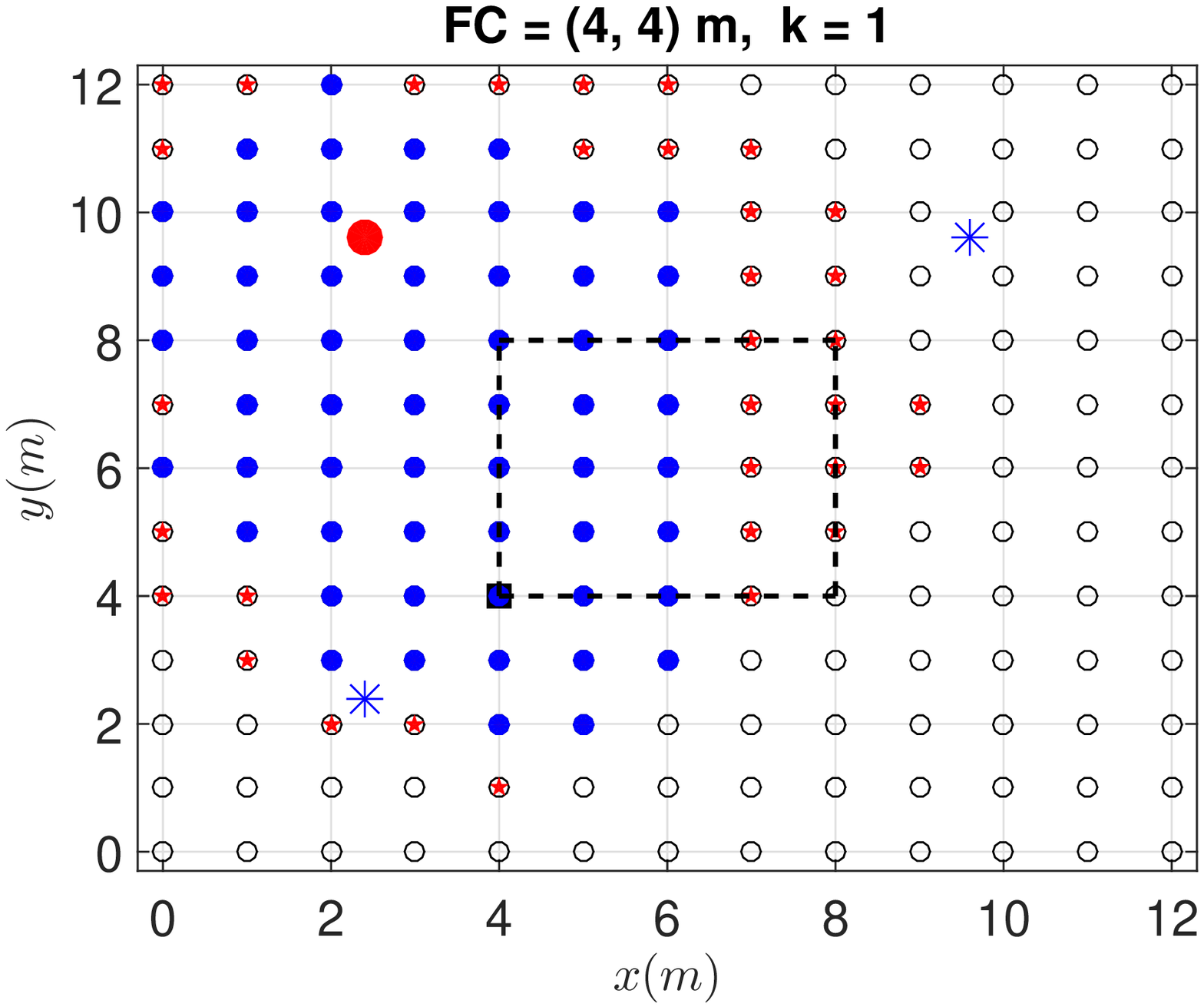}}
      \hspace*{-2em}
      \vspace{-0.1cm}
  \end{subfigure}
  \begin{subfigure}
      {\includegraphics[width=0.267\textwidth]{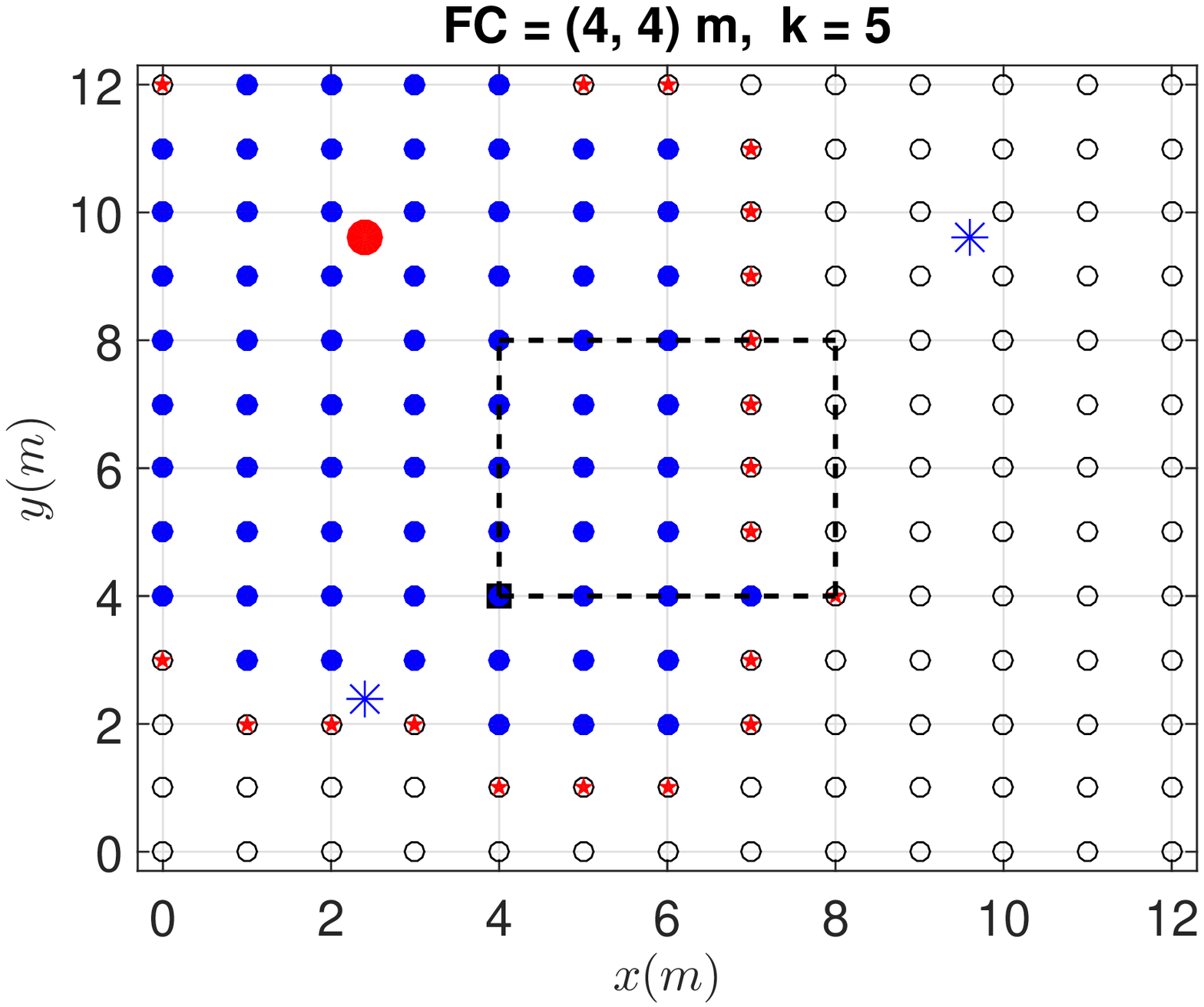}}
      \hspace*{-2em}
      \vspace{-0.1cm}
  \end{subfigure}\\
   \begin{subfigure}
      {\includegraphics[width=0.5\textwidth]{greedy_selection_legend.eps}}
      \vspace{-0.5cm}
  \end{subfigure}
\caption{An illustration of the sensor selection based on the proposed greedy algorithm for the moving FC.}
\label{fig:greedy_selection_movingFC}
\vspace{-0.3cm}
\end{figure*}
In this part, we will show the advantage of the greedy algorithm in a dynamic scenario with a moving FC. In practice, the FC could be moving, because usually it is regarded as a mobile user.
Fig.~\ref{fig:greedy_selection_movingFC} shows an example of greedy sensor selection for a moving FC, where the FC moves along the black dashed rectangle. The starting point is located at (4, 8) m, and at this position it takes 12 steps (9 steps for the local constraint and 3 steps for the global constraint) for the greedy algorithm to converge to a feasible informative set. The changing trend of the previous 11 steps is similar to  Fig.~\ref{fig:greedy_local_global}, so we merely show the results of the steps 1 and 12 in the left top subplot in Fig.~\ref{fig:greedy_selection_movingFC}. The FC then slowly moves to the next position (4, 6.67) m. For the second position, we use the selected microphone set from the first position to update the candidate set, and then solve (\ref{eq:greedy_optimize}). It is found that only 5 iterations (1 for the local constraint and 4 for the global constraint) are required to obtain convergence. Subsequently, the FC continues moving. For the next positions, the greedy algorithm only requires about 6 iterations to converge. Hence, in the dynamic scenario with a moving FC, the proposed greedy approach can significantly save computational resources.

An interesting phenomenon occurs in Fig.~\ref{fig:greedy_selection_movingFC}. As the FC moves further away form the source, we can clearly see the importance of the sensors that are close to the interference. When the FC is located at (4, 6.67) m, two sensors close to the interference are also selected. This cannot be distinguished when FC = (4, 8) m, where the FC is closer to the source. Hence, we can conclude that the sensors that are close to the source, to the FC and to the interference are informative, and they are helpful to enhance the target source, to save transmission costs and to cancel the interfering sources, respectively.

\subsection{Complexity analysis}\label{sec:exp:complexity}
\begin{figure}
  \centering
  \vspace{-0.3cm}
  \includegraphics[width=0.45\textwidth]{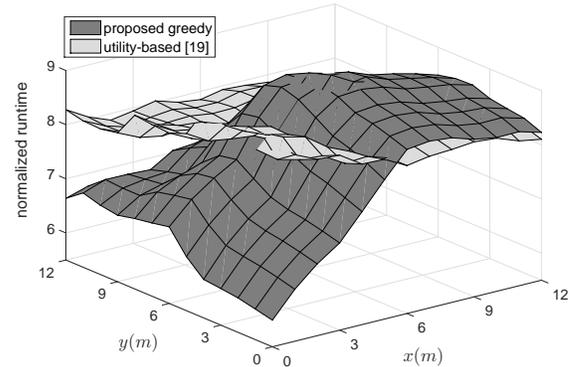}\\
  \vspace{-0.4cm}
  \caption{Execution time in terms of different initial points.}
  \label{fig:greedy_runtime}
  \vspace{-0.3cm}
\end{figure}

In this subsection, we will compare the computational complexity of the greedy algorithms to that of the model-driven approach. For the model-driven approach, its complexity is of the order of $\mathcal{O}(M^3)$, so we use $M^3$ in the worst case for analysis without loss of generality. For the proposed greedy algorithm (i.e., Algorithm 1), suppose that $J$ iterations are required to converge, in each iteration its complexity is of the order of $\mathcal{O}(|\mathcal{S}_1|^3)$, thus we can use $\sum_{j=1}^{J}|\mathcal{S}_1|^3$ to represent its computational complexity. For the utility-based greedy algorithm (i.e., Algorithm 2), we can find that its computational complexity is of the order of $\mathcal{O}(|\mathcal{S}_2|^2(|\mathcal{S}_1|-|\mathcal{S}_2|))$ for each iteration from~[19], thus $\sum_{j=1}^{J}|\mathcal{S}_2|^2(|\mathcal{S}_1|-|\mathcal{S}_2|)$ can be exploited to represent its total complexity.

Fig.~\ref{fig:greedy_runtime} compares the execution time of the two aforementioned greedy strategies. The execution time is normalized by the runtime of model-driven method, whose runtime is 1 as benchmark. From Fig.~\ref{fig:greedy_runtime}, we can see that the execution time of the proposed greedy algorithm depends on the initial point $z_0$, as it will be more expensive for the initial points that are further from the target source. Furthermore, for most initial points the proposed algorithm is computationally more efficient than the utility-based method, because we need much less iterations (20 iterations compared to 90 iterations approximately).

Although the computational complexity of the greedy algorithms could be larger than that of the model-driven algorithm, it belongs to the data-driven schemes. That is, we do not need to know the number of microphones in an environment, and it is unnecessary to inform all microphones to transmit their recorded data to the FC to estimate the statistics beforehand. Instead, it is only required to include the closest neighboring microphone nodes gradually, the FC then updates the statistics and decides the informative subset. Hence, compared to the model-driven method which is suitable for static environments, the greedy approach can be applied to dynamic scenarios, especially with infinite candidate microphones.

\section{Conclusions}\label{sec:conclusion}
In this work, we considered selecting the most informative microphone subset for the MVDR beamfomer based noise reduction. The proposed strategies were formulated through minimizing the transmission cost with the constraint on noise reduction performance. Firstly, if the statistics (e.g., the estimates of noise correlation matrices) are available, the microphone subset selection can be solved in a model-driven scheme by utilizing the convex optimization techniques. Additionally, in order to make the sensor selection capable of dynamic environments, a greedy approach in a data-driven scheme was proposed as an extension of the model-driven method. The performance of the proposed greedy algorithm converges to that of the model-driven approach. More importantly, it works more effectively in dynamic environments (e.g., with a moving FC). We concluded that in order to enhance the speech source as well as to save transmission costs, the sensors close to the source signal, those close to the FC and some close to the interferences are of larger probability to be selected, and they are helpful to enhance the target source, to save transmission costs and to cancel the interfering source, respectively. In a more general WASN, the network could consist of larger number of microphone nodes, which makes the model-driven approach  impractical. The greedy algorithm is still a possible alternative to handle the microphone subset selection problem.

\ifCLASSOPTIONcaptionsoff
  \newpage
\fi


\end{document}